\documentclass[showpacs,aps,prd,preprint,nofootinbib,showkeys,unsortedaddress,raggedbottom]{revtex4-1}
\pdfoutput=1

\usepackage{bm}
\usepackage{amsmath}
\usepackage{graphicx}
\usepackage{subfigure}
\usepackage{float}
\usepackage[usenames,dvipsnames]{color}
\definecolor{darkblue}{RGB}{0,0,196}
\usepackage[colorlinks=true,linkcolor=darkblue,citecolor=darkblue,urlcolor=darkblue]{hyperref}

\usepackage{setspace}
\usepackage{footmisc}
\usepackage[makeroom]{cancel}

\newcommand{\checked}[1]{}

\def\be{\begin{equation}}
\def\ee{\end{equation}}
\def\ba{\begin{eqnarray}}
\def\ea{\end{eqnarray}}

\begin{document}

\title{Anisotropic hydrodynamics with number-conserving kernels}

\author{Dekrayat Almaalol}
\affiliation{Department of Physics, Kent State University, Kent, OH 44242 United States}

\author{Mubarak Alqahtani} 
\affiliation{Department of Basic Sciences, College of Education, Imam Abdulrahman Bin Faisal University, Dammam 34212, Saudi Arabia.}

\author{Michael Strickland}
\affiliation{Department of Physics, Kent State University, Kent, OH 44242 United States}


\begin{abstract}
We compare anisotropic hydrodynamics (aHydro) results obtained using the relaxation-time approximation (RTA) and leading-order (LO) scalar $\lambda \phi^4$ collisional kernels.  We extend previous work by explicitly enforcing number conservation through the incorporation of a dynamical chemical potential (fugacity) in the underlying aHydro distribution function.  We focus on the case of a transversally homogenous and boost-invariant system obeying classical statistics and compare the relevant moments of the two collisional kernels.  We then compare the time evolution of the aHydro microscopic parameters and components of the energy-momentum tensor.  We also determine the non-equilibrium attractor using both the RTA and LO conformal $\lambda \phi^4$ number-conserving kernels.  We find that the aHydro dynamics receives quantitatively important corrections when enforcing number conservation, however, the aHydro attractor itself is not modified substantially.
\end{abstract}




\maketitle

\section{Introduction}
\label{sec:intro}

In the kinetic theory, the collisional kernel provides the microscopic input to the Boltzmann equation and encodes the dynamical processes which drive the system toward equilibrium \cite{deGroot}.  In hydrodynamics approaches which are based on kinetic theory, moments of the collisional kernel are used and, therefore, the choice of a specific collisional kernel dictates the manner in which the resulting fluid description approaches equilibrium.  In the anisotropic hydrodynamics (aHydro) framework \cite{Florkowski:2010cf,Martinez:2010sc,Alqahtani:2017mhy}, for example, most papers to date have used the relaxation-time approximation (RTA) for the collisional kernel~\cite{anderson1974relativistic}.  Despite its simplicity, 3+1d aHydro codes which use the RTA do a quite reasonable job in describing experimental observations of identified hadron spectra, elliptic flow, Hanbury-Brown-Twiss radii, etc. \cite{Alqahtani:2017jwl,Alqahtani:2017tnq,Almaalol:2018gjh}.  Given this success, it is desirable to make the underlying aHydro equations of motion more realistic by using collisional kernels associated with an actual quantum field theory.  Of course, the eventual goal is to use realistic scattering kernels based on quantum chromodynamics \cite{forth}.  Herein, we take a small step in this direction by making comparisons between results obtained using the RTA and leading-order (LO) scalar $\lambda \phi^4$ collisional kernels.

In our previous work~\cite{Almaalol:2018ynx}, we demonstrated how to use a general $2 \leftrightarrow 2$ collisional kernel in the aHydro formalism and then specialized to the case of a LO scalar $\lambda \phi^4$ theory.  We applied the aHydro equations to a 0+1d conformal system undergoing boost-invariant longitudinal expansion. Our results demonstrated that the system dynamically produced higher anisotropy when using the LO scalar kernel than when using the RTA kernel. We also demonstrated that the system approached its non-equilibrium attractor more slowly with the LO scalar kernel.  

In this work, we extend the analysis presented in Ref.~\cite{Almaalol:2018ynx} by enforcing number conservation using both the RTA and LO massless $\lambda \phi^4$ kernels.  In both cases, we generalize the Romatschke-Strickland form \cite{Romatschke:2003ms,Romatschke:2004jh} to include a dynamical chemical potential.  We derive the necessary aHydro equations of motion using the 0$^{\rm th}$, 1$^{\rm st}$, and 2$^{\rm nd}$ moments of the Boltzmann equation, solve the resulting ordinary differential equations numerically, and discuss the effect of enforcing number conservation with both the RTA and LO scalar kernels.  Using the resulting equations of motion, we also determine the differential equation obeyed by the aHydro dynamical ``attractor'' \cite{Strickland:2017kux,Almaalol:2018ynx}, now taking into account number conservation.  The attractor drives the early-time dynamical evolution of the system and is important in understanding the hydrodynamization of the quark-gluon plasma \cite{Heller:2015dha,Keegan:2015avk,Florkowski:2017olj,Romatschke:2017vte,Bemfica:2017wps,Spalinski:2017mel,Romatschke:2017acs,Behtash:2017wqg,Florkowski:2017jnz,Florkowski:2017ovw,Denicol:2018pak}.
 
 The structure of the paper is as follows. We present the setup in Sec.~\ref{sec:setup}.  In Sec.~\ref{sec:kernelchem} we introduce the RTA and LO scalar collisional kernels, taking into account a finite chemical potential. In Sec.~\ref{sec:aHydroeqns}, the aHydro equations are presented for a number conserving theory.  In Sec.~\ref{sec:kermom} we compute the necessary moments using both collisional kernels. In Sec.~\ref{sec:numericalsolution} we present representative numerical solutions of the aHydro equations of motion, comparing the LO scalar collisional kernel and the RTA collisional kernel with and without number conservation.  In this section, we also present the aHydro non-equilibrium dynamical attractor and compare to previously obtained results.  In Sec.~\ref{sec:conclusions} we provide our conclusions and an outlook for the future.
 

\section*{Conventions and notation}
\label{sec:notation}

The Minkowski metric tensor is taken to be $g^{\mu\nu}={\rm diag}(+,-,-,-)$.  The Lorentz-invariant integration measure is \mbox{$dP = \frac{d^3{\bf p}}{(2\pi)^3} \frac{1}{E_p}$} and four-vectors are decomposed as, e.g. $p^\mu = (E_p,{\bf p})$.  In what follows, we will work in the massless limit $m \rightarrow 0$ such that $E_p = |{\bf p}|$.

\section{Setup}
\label{sec:setup}

In our prior paper \cite{Almaalol:2018ynx}, we compared the equations of motion, pressure anisotropies, attactor, etc. resulting from the use of a $2 \leftrightarrow 2$ scalar collisional kernel and the Anderson-Witting kernel (relaxation time approximation or RTA) \cite{anderson1974relativistic}.  In that work, we did not explicitly take into account number conservation in the scalar theory nor did we enforce it in the RTA equations of motion.  In order to accomplish this, we generalize the distribution function ansatz to include a finite chemical potential and then use the zeroth moment of the Boltzmann equation to provide the additional equation of motion required.  We will perform our analysis for a transversally homogeneous and boost-invariant system (0+1d) in which case it suffices to introduce one anisotropy parameter \cite{Martinez:2012tu,Alqahtani:2017mhy}.  In particular, we consider the Romatschke-Strickland form for massless particles with a chemical potential and classical statistics~\cite{Martinez:2012tu,Bazow:2013ifa,Molnar:2016gwq}
\ba
f_p &=& \exp\!\left(-\frac{1}{\Lambda} \sqrt{p_\perp^2+(1+\xi) ({\bf p}\cdot\hat{\bf n})^2} + \frac{\mu}{\Lambda} \right) \, ,
\nonumber\\
&=& \gamma f_p^0 \, ,
\label{eq:frs1}
\ea
\checked{md}
where
$\gamma \equiv \exp\left( {\mu}/{\Lambda}\right) $ is the particle fugacity and
\be
f_p^0 \equiv \exp\!\left(-\frac{1}{\Lambda} \sqrt{p_\perp^2+(1+\xi) ({\bf p}\cdot\hat{\bf n})^2}\right) .
\ee
\checked{md}
is the zero chemical potential distribution function.  In the above expressions, $\xi$ is the anisotropy parameter ($-1 < \xi < \infty$), $\Lambda$ is the transverse temperature, and $\hat{\bf n}$ is a unit vector along the anisotropy direction, which is typically taken to be the beamline direction, i.e. $ \hat{\bf n}  =  \hat{\bf z}$.  Both $\xi$ and $\Lambda$ depend on spacetime in general, but we suppress this dependence for compactness of the notation.   

\section{Collisional kernels at finite chemical potential}
\label{sec:kernelchem}

In this section, we present the modifications necessary to extend our prior analyses of both the scalar and RTA collisional kernels to finite chemical potential.  We will use the Boltzmann equation to obtain the necessary aHydro equations of motion
\be
p^\mu \partial_\mu f_p =  C[f_p] \, ,
\label{eq:boltzmann1}
\ee
\checked{md}
where $f_p = f({\bf p})$ is the one-particle distribution function and the collisional kernel $C[f_p]$ is a functional which encodes the details of the specific microscopic interactions.

\subsection{Scalar collisional kernel at finite chemical potential}
\label{sec:kernel}

We will consider massless scalar $\lambda \phi^4$ to leading order in the coupling.  The elastic $2 \leftrightarrow 2$ scattering kernel with classical statistics can be written in the form~\cite{Almaalol:2018ynx,Jeon:1995zm}
\ba 
C_{\rm sc}[f_p] &=&\frac{1}{32} \int dK dK' dP' \, | {\cal M} |^2 \, (2\pi)^4 \delta^{(4)}(k^\alpha + k'^\alpha - p^\alpha - p'^\alpha) \, {\cal F}(k,k',p,p') \, ,
\label{eq:scalarkernel}
\ea
\checked{md}
where
\be
{\cal F}(k,k',p,p') \equiv f_k f_{k'} - f_p f_{p'} \, ,
\label{eq:Fdef}
\ee
\checked{md}
with ${\cal M}$ being the invariant scattering amplitude.  For the case considered one has $|{\cal M}|^2 = \lambda^2$ with $\lambda$ being the scalar coupling constant. 

Using Eq.~(\ref{eq:frs1}) one can see immediately that the distribution function factorizes
\ba 
{\cal F}(k,k',p,p') = \gamma^2 {\cal F}^0(k,k',p,p') \, ,
\label{eq:Fdef2}
\ea
\checked{md}
where the superscript 0 indicates the statistical factors at zero chemical potential.  From this, it follows that
\be
C_{\rm sc}[f_p] = \gamma^2 C_{\rm sc}[f_p^0] \, ,
\label{eq:sccases}
\ee
\checked{md}
where the subscript `sc' indicates `scalar'.

%
\subsection{RTA kernel at finite chemical potential}

At finite chemical potential, the RTA collisional kernel can be written as
\ba 
C_{\rm RTA}[f_p] &=& \frac{E_p}{\tau_{\rm eq}} \left[ f_{\rm eq} - f_p \right] \, 
\ea 
\checked{m}
where
\be
f_{\rm eq}(p/T) \equiv \Gamma \exp(-|{\bf p}|/T) = \Gamma f_{\rm eq}^0 \, ,
\ee
\checked{md}
with $T$ being the effective temperature and $\Gamma$ being the effective fugacity.  Above $\tau_{\rm eq} = 5 \bar\eta/T$ with $\bar\eta \equiv \eta/s$ being the specific shear viscosity~\cite{Denicol:2010xn,Denicol:2011fa}.  As a result, one has
\be
C_{\rm RTA}[f_p] = \frac{E_p}{\tau_{\rm eq}} \left[ \Gamma f_{\rm eq}^0 - \gamma f_p^0 \right] .
\label{eq:rtamu1}
\ee
\checked{md}
In order to fix the effective temperature and fugacity we require the right hand sides of the zeroth and first moments of the Boltzmann equation to vanish.  These constraints enforce number and energy-momentum conservation, respectively.  They result in the following two relations
\ba
T &=& {\cal R}(\xi) \sqrt{1+\xi} \, \Lambda \, , \label{eq:updmatch1} \\
\Gamma &=& \frac{\gamma}{(1+\xi)^2 {\cal R}^3(\xi)} \, ,
\label{eq:updmatch2} 
\ea
\checked{md}
where \cite{Martinez:2010sc}
\be
{\cal R}(\xi) = \frac{1}{2}\left[\frac{1}{1+\xi}
+\frac{\arctan\sqrt{\xi}}{\sqrt{\xi}} \right] .
\label{eq:rfunc}
\ee
\checked{md}

Using (\ref{eq:updmatch2}) we can write the RTA collisional kernel at finite chemical potential as
\be
C_{\rm RTA}[f_p] = \frac{\gamma E_p}{\tau_{\rm eq}} \left[  \frac{ f_{\rm eq}^0 }{(1+\xi)^2 {\cal R}^3(\xi)} - f_p^0 \right] .
\label{eq:rtamu2}
\ee
\checked{md}

\section{aHydro equations of motion at finite chemical potential}
\label{sec:aHydroeqns}

In this section, we derive the conformal 0+1d equations of motion using both the LO scalar and RTA collisional kernels.   The starting point is the Boltzmann equation \eqref{eq:boltzmann1} with the collisional kernel given by either \eqref{eq:scalarkernel} or \eqref{eq:rtamu1}.   As usual, in anisotropic hydrodynamics we take moments of the Boltzmann equation \cite{Alqahtani:2017mhy}.  The zeroth-moment equation is
\be
\partial_\mu n^{ \mu } = 0 \, ,
\label{eq:neq}
\ee
\checked{md}
where $n^\mu = n u^\mu$ with $n$ being the number density.  The right hand side of \eqref{eq:neq} vanishes automatically for the scalar collisional kernel and vanishes in RTA due to the matching conditions \eqref{eq:updmatch1} and \eqref{eq:updmatch2}. Using \eqref{eq:frs1} one has $n = \gamma n^0_{\rm eq}(\Lambda)/\sqrt{1+\xi}$, where $n_{\rm eq}^0$ is the equilibrium number density at zero chemical potential.  As a result, the zeroth moment equation becomes
\be 
\partial_{\tau} \ln \gamma+ 3 \,\partial_{\tau} \ln \Lambda - {\frac{1}{2}}\frac{\partial_{\tau} \xi}{1+\xi} + \frac{1}{\tau} =0 \, .
\label{eq:final0thmom}
\ee
\checked{{md}}
The first-moment equation encodes energy-momentum conservation
\be
\partial_\mu T^{ \mu \nu } = 0 \, ,
\ee
\checked{md}
where, once again, the right hand side vanishes automatically for the scalar collisional kernel and vanishes in RTA due to the matching conditions \eqref{eq:updmatch1} and \eqref{eq:updmatch2}.  Expanding the first moment equation using \eqref{eq:frs1}, one obtains
\be 
\partial_{\tau} \ln \gamma + 4 \,\partial_{\tau} \ln \Lambda + \frac{ {\cal R}^\prime(\xi)}{\cal R (\xi)} \partial_{\tau} \xi= \frac{1}{\tau} \left[\frac{1}{\xi(1+\xi){\cal R (\xi)}} - \frac{1}{\xi} -1 \right] .
\label{eq:final1stmom}
\ee
\checked{md}

Finally, we need one equation from the second moment which is obtained by taking the $zz$-projection minus one third of the sum of the $xx$, $yy$, and $zz$ projections \cite{Tinti:2013vba}.  For a general collisional kernel, one obtains 
\be
\frac{1}{1+\xi} \partial_\tau \xi  - \frac{2}{\tau} = {\cal K}  \, ,
\label{eq:fin2ndmom}
\ee
\checked{md}
with
\be
{\cal K} \equiv \frac{{\cal C}^{xx}}{I_{x}} -  \frac{{\cal C}^{zz}}{I_{z}}  = \frac{\pi^2\Lambda}{4\gamma} \left[ (1+\xi)^{1/2} \bar{\cal C}^{xx}(\xi)  - (1+\xi)^{3/2} \bar{\cal C}^{zz}(\xi) \right],
\label{eq:finCgensimp}
\ee
\checked{md}
where 
\be
\bar{\cal C}^{\mu\nu} \equiv \frac{1}{\Lambda^6} \int dP \,  p^\mu p^\nu \, C[f_p] \, ,
\label{eq:calcdef}
\ee
\checked{md}
and
\be
I_i \equiv \int \! \frac{d^3p}{(2\pi)^3} \, p_i^2 f_p = \gamma I^0_i\, .
\ee
\checked{md}

\section{Moments of the collisional kernels}
\label{sec:kermom}

In order to proceed, we need to compute ${\cal K}$ \eqref{eq:finCgensimp} using both the scalar and the RTA collisional kernels.  After some algebra, it can be shown that in RTA one has
\ba
{\cal K}_{\rm RTA} &=&  \frac{\Lambda}{5 \bar{\eta}} \xi (1+\xi)^\frac{3}{2} {\cal R}^3(\xi) \nonumber \\
&=& \frac{1}{\tau_{\rm eq}}  \xi (1+\xi)  {\cal R}^2(\xi)\, .
\ea
\checked{md}
In order to compare the scalar case to RTA it is convenient to pull out the overall factor of $\lambda^2$ by defining $\tilde{\cal C}^{ii} = \bar{\cal C}^{ii}/\lambda^2$, which gives
\be
{\cal K}_{\rm sc} =  \frac{\pi^2 \lambda^2 \Lambda }{4\gamma} \left[ (1+\xi)^{1/2} \tilde{\cal C}^{xx}_{\rm sc} (\xi)  - (1+\xi)^{3/2} \tilde{\cal C}^{zz}_{\rm sc} (\xi) \right],
\label{eq:finCgensimp2}
\ee
\checked{md}
For the scalar collisional kernel we must evaluate the remaining 8-dimensional integrals $\tilde{\cal C}^{xx}(\xi)$ and $\tilde{\cal C}^{zz}(\xi)$ numerically  \cite{gsl}.  

Additionally, if we want to make a proper comparison between dynamics subject to the RTA and scalar collisional kernels, we should match the two collisional kernels in the near equilibrium limit.  In order to do this, we expand both results to leading order in $\xi$ and match the leading-order coefficients.  This can be done with the full ${\cal K}$ function or using either term contributing to  $\cal K$.  Following our previous paper, we evaluate $\bar{\cal C}^{zz}(\xi)$ for both collisional kernels and equate the leading-order coefficients~\cite{Almaalol:2018ynx}.\footnote{Once the matching is done using $\bar{\cal C}^{zz}(\xi)$, it is guaranteed to work for $\bar{\cal C}^{xx}(\xi)$ and hence ${\cal K}$.}

For the RTA kernel, the small-$\xi$ expansion can be done analytically with the result being
\be
\lim_{\xi \rightarrow 0} \bar{\cal C}^{zz}_{\rm RTA}= \frac{8 \gamma}{15 \pi^2 \bar\eta} \xi + {\cal O}(\xi^2) \, .
\ee
\checked{md}
For the scalar kernel, the numerical result is
\be
\lim_{\xi \rightarrow 0} \bar{\cal C}^{zz}_{\rm sc}  = \alpha \gamma^2 \lambda^2 \xi + {\cal O}(\xi^2) \, ,
\label{eq:smallxi3}
\ee
\checked{md}
with $\alpha \simeq 0.4394 \pm 0.0002$~\cite{Almaalol:2018ynx}.  

Equating the leading-order RTA and scalar kernel results listed above, we obtain the following matching condition
\be
  \lambda^2  = \frac{8}{15 \pi^2  \alpha  \gamma  \bar\eta } \, .
\label{eq:match1}
\ee 
\checked{md}
With this, Eq.~\eqref{eq:finCgensimp2} becomes
\be
{\cal K}_{\rm sc} = \frac{2\Lambda}{15  \alpha  \gamma^2  \bar\eta } \left[ (1+\xi)^{1/2} \tilde{\cal C}^{xx}_{\rm sc} (\xi)  - (1+\xi)^{3/2} \tilde{\cal C}^{zz}_{\rm sc} (\xi) \right] ,
\label{eq:finCgensimp3}
\ee
\checked{md}

\subsection{Final second moment equations}

Using the matching condition \eqref{eq:match1}, one can write the second moment equation \eqref{eq:fin2ndmom} in the following compact form \cite{Strickland:2017kux,Almaalol:2018ynx}:
\be
\partial_\tau \xi  - \frac{2(1+\xi)}{\tau} + \frac{{\cal W}(\xi)}{\tau_{\rm eq}}  = 0 \, .
\label{eq:2ndmoment}
\ee
\checked{md}
For the RTA kernel, the ${\cal W}$ function is given by
\be
{\cal W}_{\rm RTA}(\xi) = \xi (1+\xi)^2 \, {\cal R}^{2}(\xi) \, ,
\label{eq:calwrta}
\ee
\checked{md}
and for the scalar collisional kernel it is
\be
{\cal W}_{sc}(\xi) \equiv \frac{2}{3 \alpha {\cal R}(\xi) }  \left[  (1+\xi)^2 \tilde{\cal C}^{zz}_{sc,0}(\xi) - (1+\xi)\tilde{\cal C}^{xx}_{sc,0}(\xi) \right].
\label{eq:calwsc}
\ee
\checked{md}

\subsection{Connection to second-order viscous hydrodynamics and the attractor}
\label{sec:attractorvars}

Based on the results contained in Ref.~\cite{Strickland:2017kux} and \cite{Almaalol:2018ynx}, once we have cast the second moment equation the form \eqref{eq:2ndmoment}, the second-moment equation and associated attractor equation can then be written in terms of the shear viscous correction, $\Pi$.  Using
\be
\overline\Pi(\xi) \equiv \frac{\Pi}{\epsilon} = \frac{1}{3} \left[ 1 - \frac{{\cal R}_L(\xi)}{\cal R(\xi)} \right]  .
\label{eq:pixirel}
\ee 
\checked{m}
one obtains
\be
\frac{\dot\Pi}{\epsilon} + \frac{\Pi}{\epsilon\tau} \left( \frac{4}{3} - \frac{\Pi}{\epsilon}  \right) - \frac{2(1+\xi)\overline\Pi'(\xi)}{\tau} + \frac{{\cal W}(\xi) }{\tau_{\rm eq}} {\overline\Pi}^\prime(\xi) = 0\, .
\label{eq:2ndmomf2}
\ee
\checked{m}
where it is understood that $\xi = \xi(\overline\Pi)$ with $\xi(\overline\Pi)$ being the inverse function of $\overline\Pi(\xi)$.  For details concerning construction of this inverse function, we refer the reader to Ref.~\cite{Strickland:2017kux}.

Transforming to ``attractor variables''
\ba
w &\equiv& \tau T(\tau) \, , \nonumber \\
\varphi &\equiv& \tau \frac{\dot w}{w} \, ,
\label{wdot1}
\ea
\checked{m}
one obtains the following first-order differential equation by combining the first moment with Eq.~\eqref{eq:2ndmomf2}~\cite{Strickland:2017kux}
\be
\overline{w} {\mathcal \varphi} \frac{\partial \varphi}{\partial \overline{w}}  =  \left[ \frac{1}{2} (1+\xi)  - \frac{\overline{w}}{4} {\cal W} \right] \overline\Pi' \, ,
\label{eq:ahydroattractoreq2}
\ee
\checked{m}
where $\overline{w} \equiv w/c_\pi$ with $c_\pi = 5 \bar\eta$.  Once the ``amplitude'' $\varphi$ is determined by solving \eqref{eq:ahydroattractoreq2} subject to the appropriate boundary condition at $\overline{w}=0$, one can obtain the pressure anisotropy using
\be
\frac{{\cal P}_L}{{\cal P}_T} = \frac{3 - 4 \varphi}{2 \varphi - 1} \, .
\ee
\checked{m}

\begin{table}[t!]
$
\begin{array}{|c|c|}
\hline
\text{\hspace{1.7cm}} & \text{\bf result}  \\
\hline
 c_0 & 0 \\
 c_1 & 1 \\
 c_2 & 0.60658 \\
 c_3 & -0.068866 \\
 c_4 & 0.0077844 \\
 c_5 & -0.00062427 \\
 c_6 & 0.000034979 \\
 c_7 & -1.393\times 10^{-6} \\
\hline
\end{array}
\hspace{8mm}
\begin{array}{|c|c|}
\hline
\text{\hspace{1.7cm}} & \text{\bf result} \\
\hline
 c_8 & 4.0055\times 10^{-8} \\
 c_9 & -8.3865\times 10^{-10} \\
 c_{10} & 1.2781\times 10^{-11} \\
 c_{11} & -1.4017\times 10^{-13} \\
 c_{12} & 1.0771\times 10^{-15} \\
 c_{13} & -5.5029\times 10^{-18} \\
 c_{14} & 1.6784\times 10^{-20} \\
 c_{15} & -2.3126\times 10^{-23} \\
 \hline
\end{array}
$
\caption{Polynomial fit coefficients for the classical LO scalar ${\cal W}_{sc}(\xi)$ function defined in Eq.~(\ref{eq:calwsc}).  The fit was made assuming ${\cal W}_{sc}(\xi) = \sum_n c_n \xi^n$ and using 101 points in the range $-0.68 \leq \xi \leq 99$.}
\label{table:coeffs}
\end{table}

\section{Results}
\label{sec:numericalsolution}

We now turn to our results.  We will compare results obtained from our prior work \cite{Almaalol:2018ynx} which assumed $\mu=0$ ($\gamma=1$) using both the RTA \eqref{eq:calwrta} and scalar \eqref{eq:calwsc} collisional kernels.  For the scalar collisional kernel we tabulated ${\cal W}_{sc}(\xi)$ using 101 points in the range \mbox{$-0.68 \leq \xi \leq 99$}.  We evaluated the eight-dimensional integrals necessary using the Monte-Carlo VEGAS algorithm \cite{Almaalol:2018ynx}.  The resulting numerical data for ${\cal W}_{sc}(\xi)$ was then fit using a 15$^\text{th}$-order polynomial ${\cal W}_{sc}(\xi) = \sum_{n=0}^{15} c_n \xi^n$.  The resulting fit coefficients are listed in Table~\ref{table:coeffs}. In addition to this polynomial fit, we performed large-$\xi$ computations and extracted the leading $\xi$-scaling of the kernel in this limit, finding that \mbox{$\lim_{\xi \rightarrow \infty} {\cal W}_{sc}(\xi) = 1.3183 \, \xi^{3/2}$}.  We used the polynomial fit for all $\xi \leq 99$ and the large-$\xi$ result for $\xi > 99$.  The resulting analytic approximations for ${\cal W}_{sc}(\xi)$ were then used as an input to Eq.~(\ref{eq:2ndmoment}).

\begin{figure*}[t!]
\centerline{
\includegraphics[width=.49\linewidth]{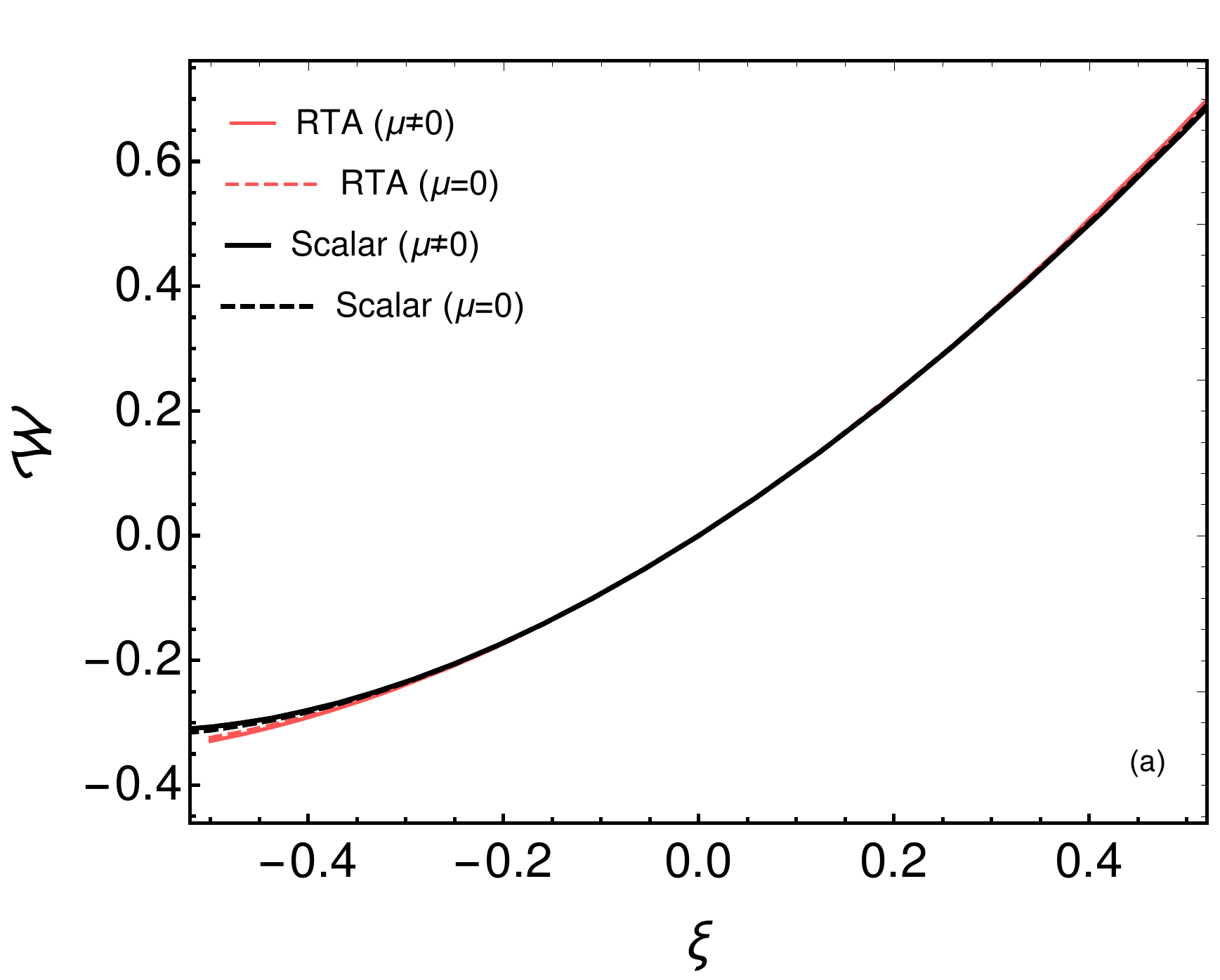}
\hspace{2mm}
\includegraphics[width=.49\linewidth]{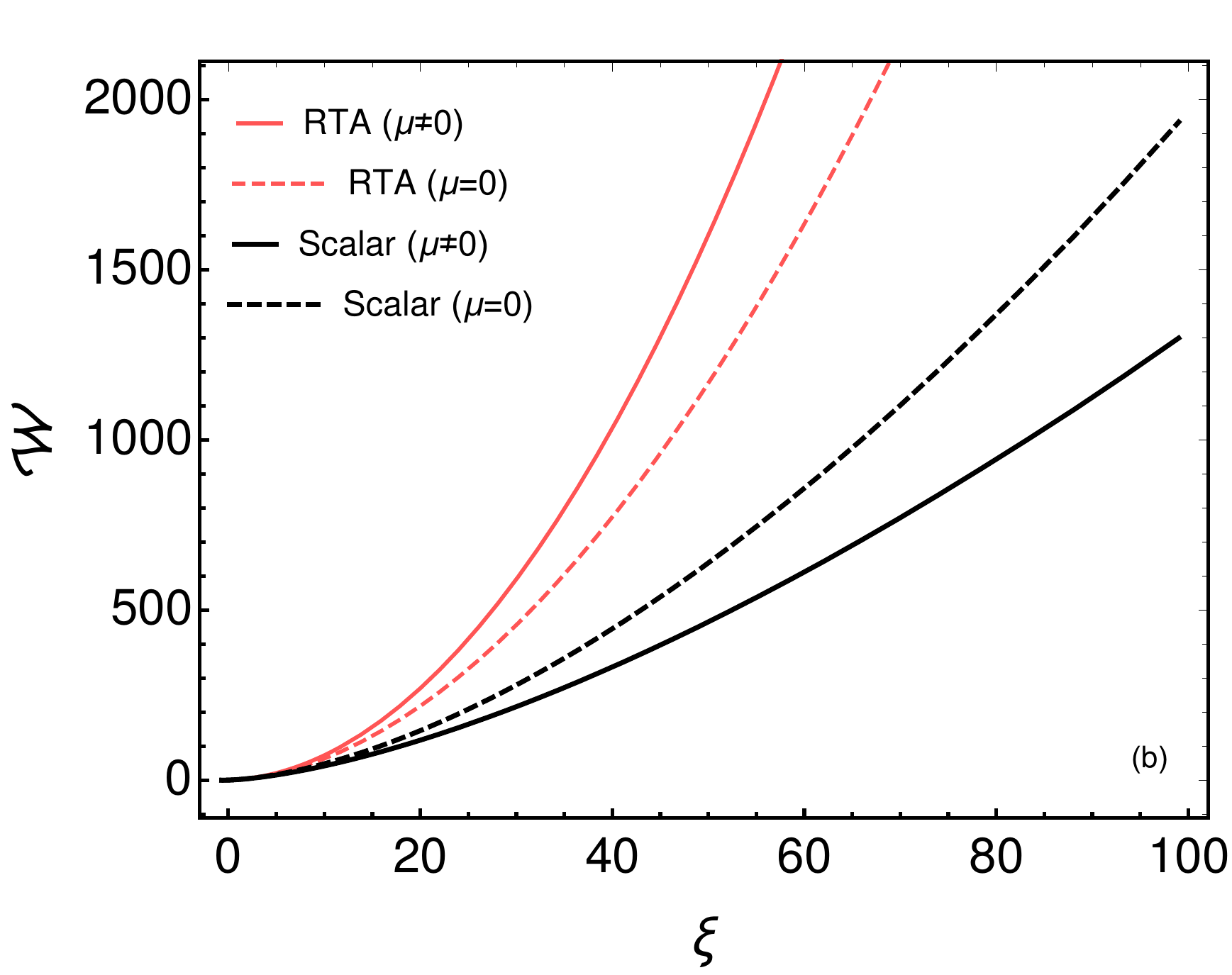}
}
\vspace{4mm}

\caption{Comparison of ${\cal W}$ from the LO scalar and RTA kernels.  Panel (a) shows the result for small values of $\xi$ and panel (b) shows the result for large values of $\xi$.  The RTA kernel results at $\mu\neq0$ and $\mu=0$ are indicated by solid red and dashed red lines, respectively.  The scalar kernel results at $\mu\neq0$ and $\mu=0$ are indicated by solid black and black dashed lines, respectively.   }
\label{fig:wplots}
\end{figure*}
\begin{figure*}[t!]
\hspace{0.5mm}
\includegraphics[width=.5\linewidth]{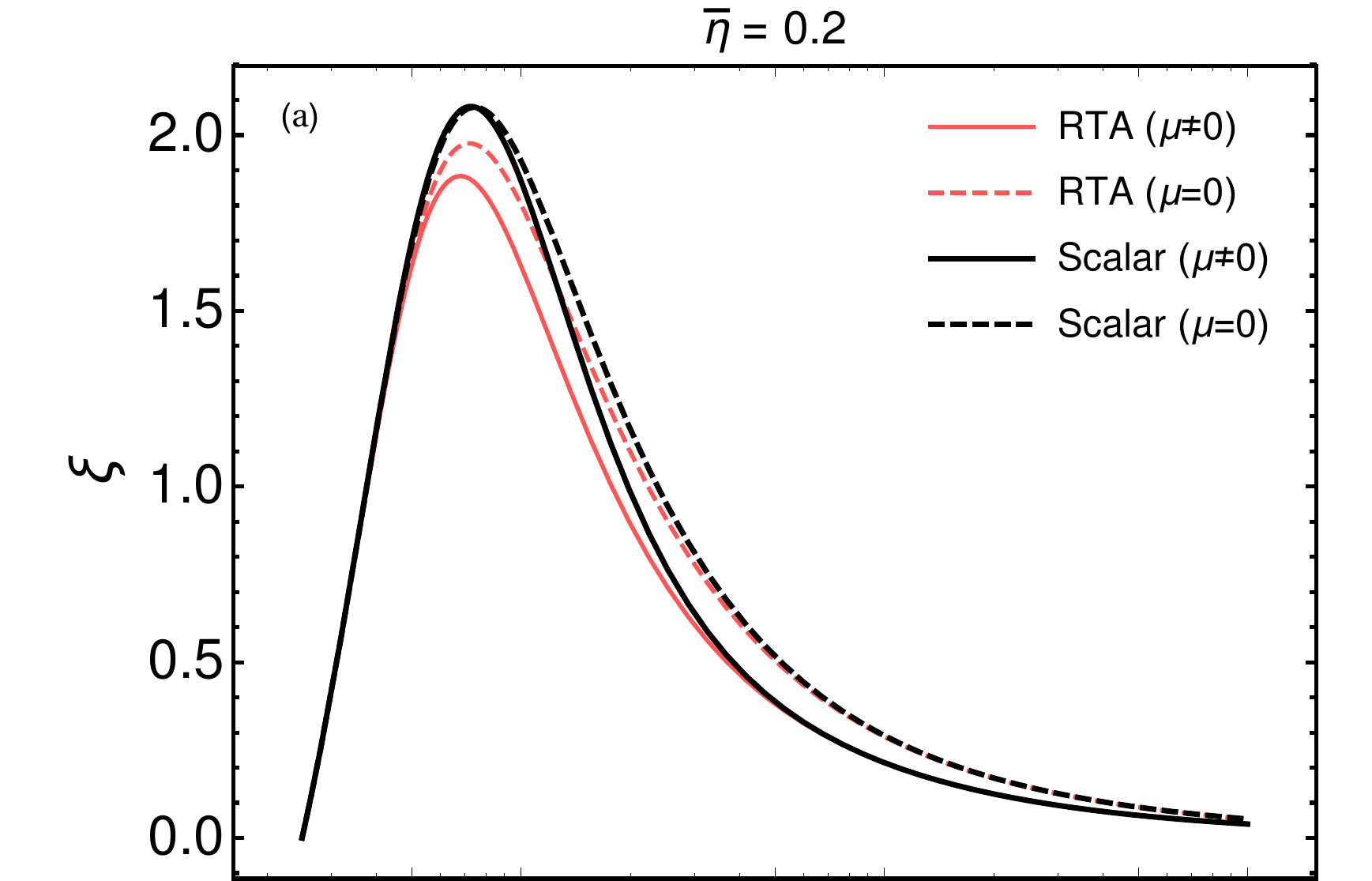}
\hspace{-7mm}
\includegraphics[width=.5\linewidth]{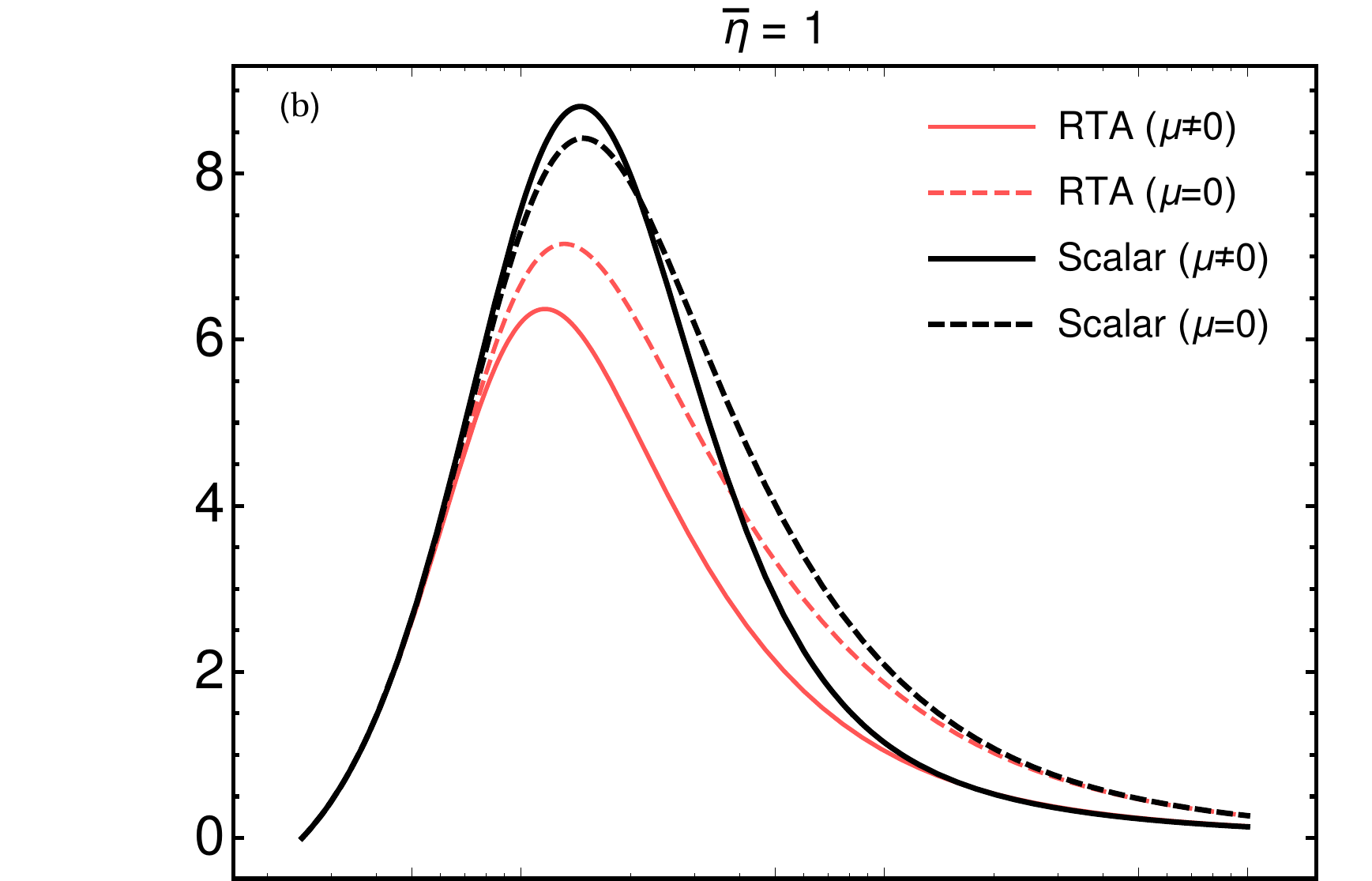}
\vspace{6mm}

\includegraphics[width=.495\linewidth]{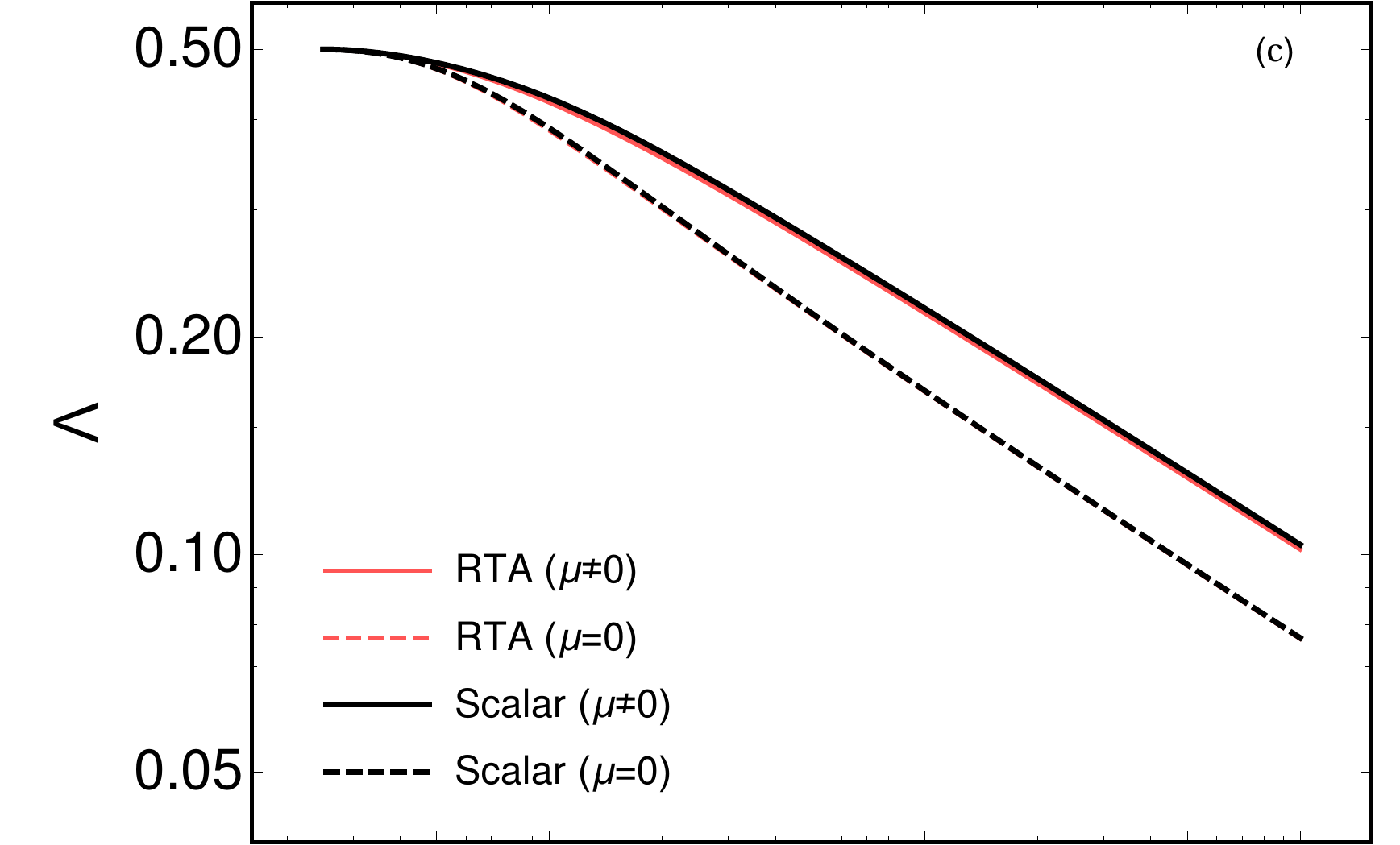}
\hspace{-7mm}
\includegraphics[width=.5\linewidth]{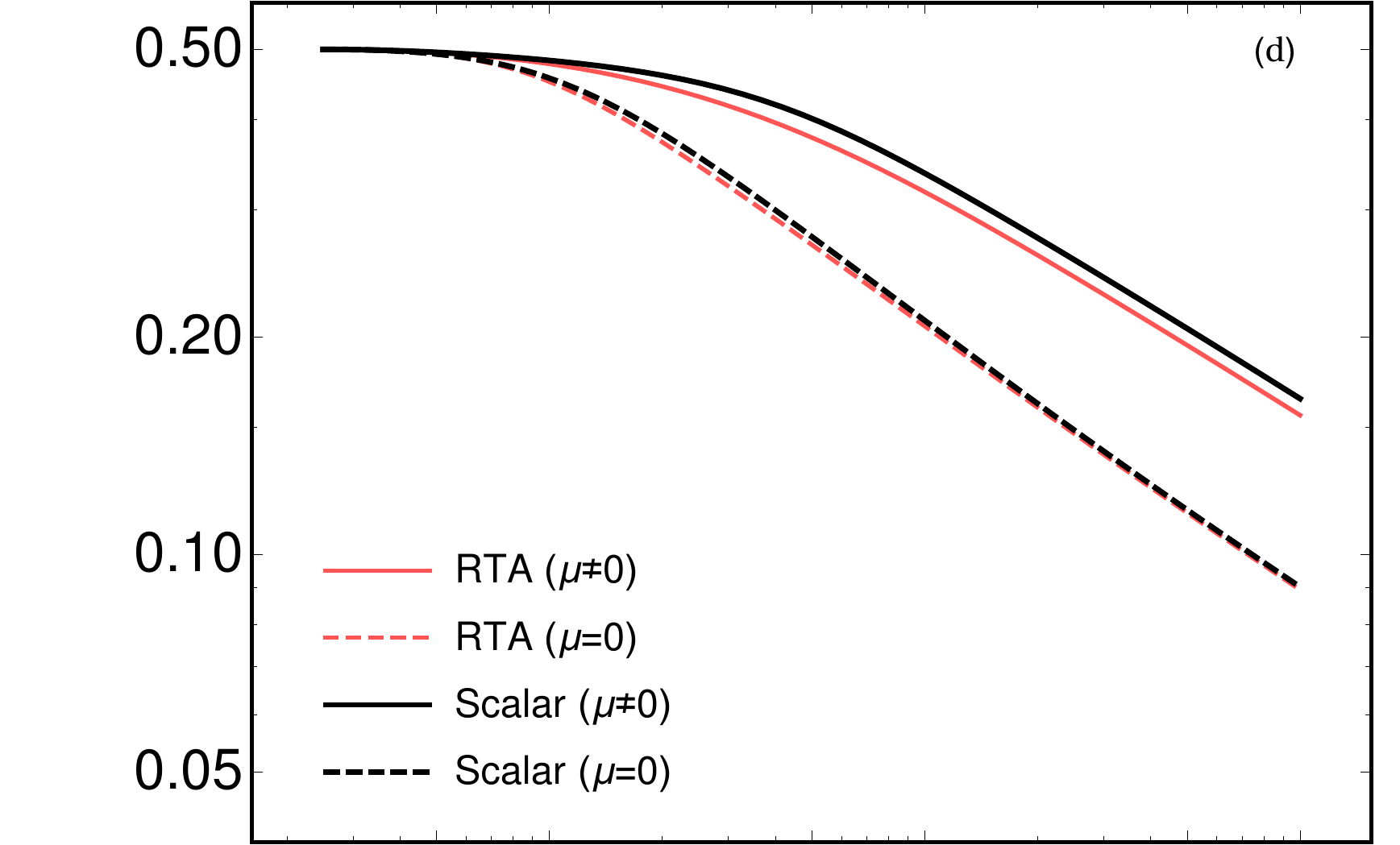}
\vspace{6mm}

\hspace{1mm}
\includegraphics[width=.46\linewidth]{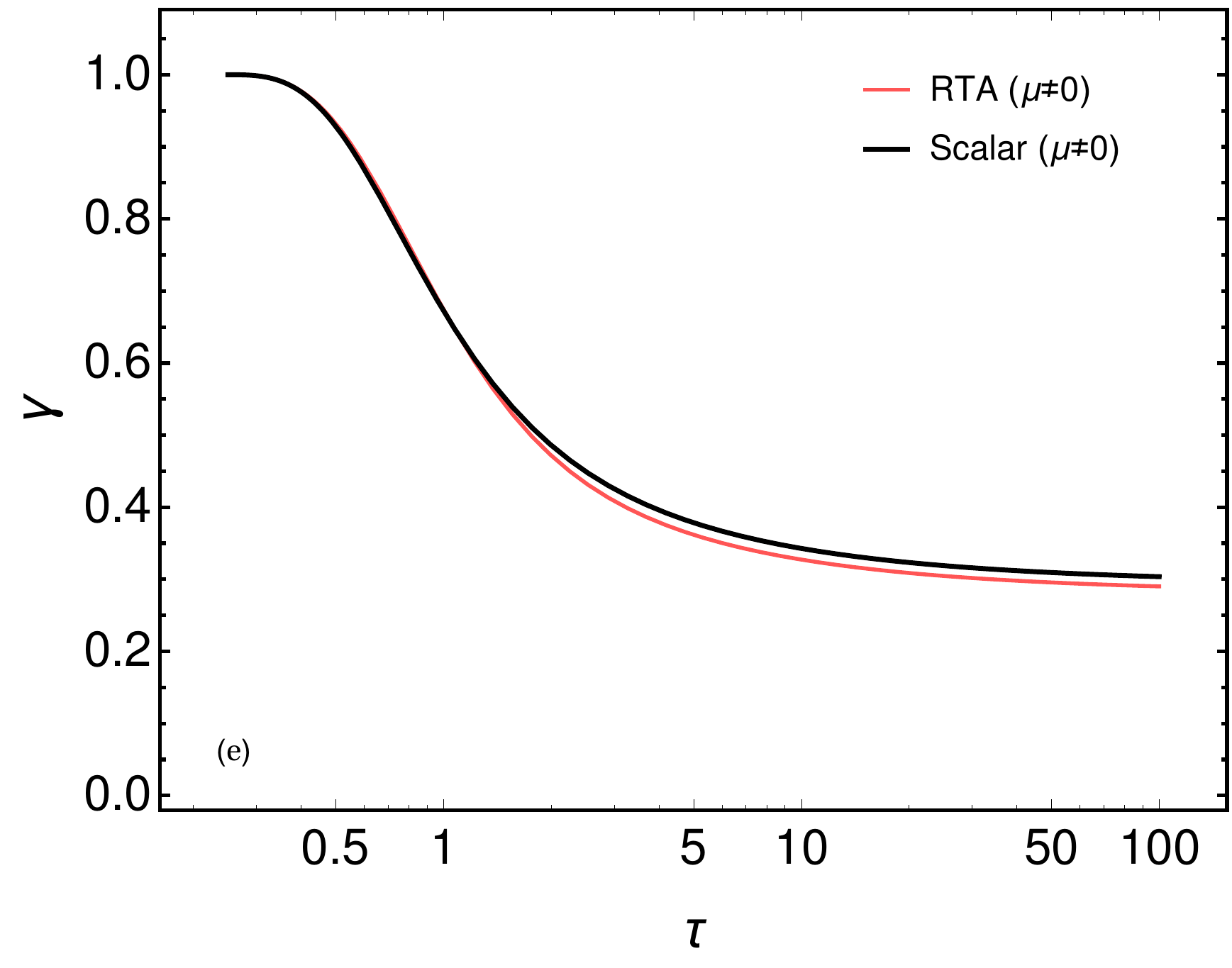}
\hspace{-1mm}
\includegraphics[width=.46\linewidth]{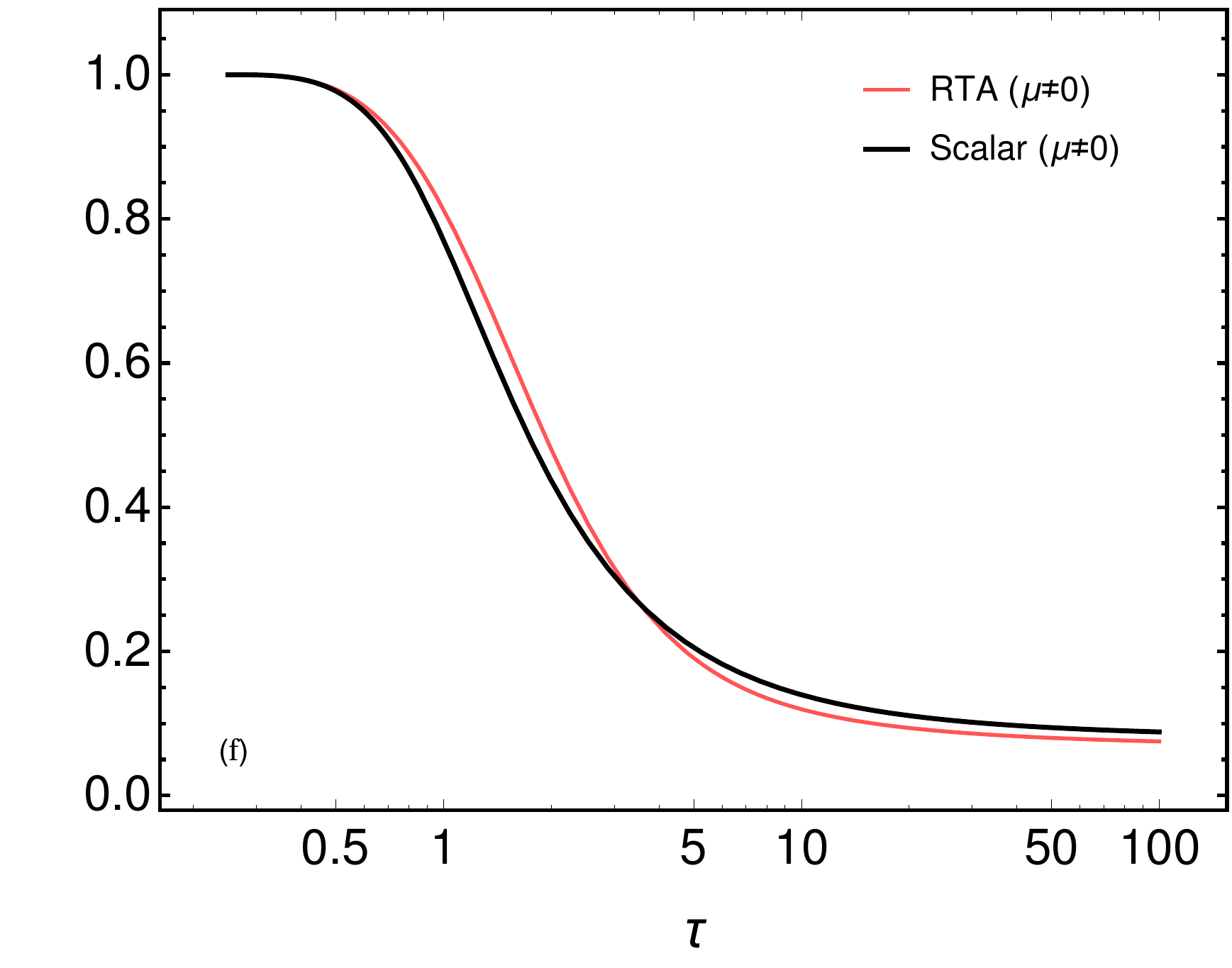}
\caption{The evolution of $\xi$ (a)-(b), the transverse temperature scale $\Lambda$ in GeV (c)-(d), and the fugacity $\gamma$ (e)-(f).  The left column panels (a), (c), and (e) show the case that $\bar\eta=0.2$ and the right column panels (b), (d), and (f) show  $\bar\eta=1$.  For this figure we assumed isotropic initial conditions with $\xi_0 = 10^{-8}$, $\tau_0= 0.25\, \rm fm/c $, $\Lambda_0 = 0.5 \,\rm GeV$, and $\gamma_0=1$.}
\label{fig:VevolPlot1}
\end{figure*}

\subsection{$\cal W$ function}

In Fig.~\ref{fig:wplots} we compare the $\cal W$ functions obtained using the LO scalar and RTA kernels.  Focussing first on the RTA kernel results (red and red dashed lines), we see that the effect of enforcing number conservation is to increase $\cal W$ at large $\xi > 0$.  As a result, one expects to see smaller momentum-space anisotropies developed when taking into account number conservation with the RTA approximation.  The scalar kernel results (black and black dashed lines) show the opposite behavior, leading to the prediction that larger momentum-space anisotropies will develop when taking into account number conservation in this case.  As we will see, this expectation is realized in our results for the early-time dynamical momentum-space anisotropy and the non-equilibrium attractor.

\begin{figure*}[t!]
\hspace{0.5mm}
\includegraphics[width=.5\linewidth]{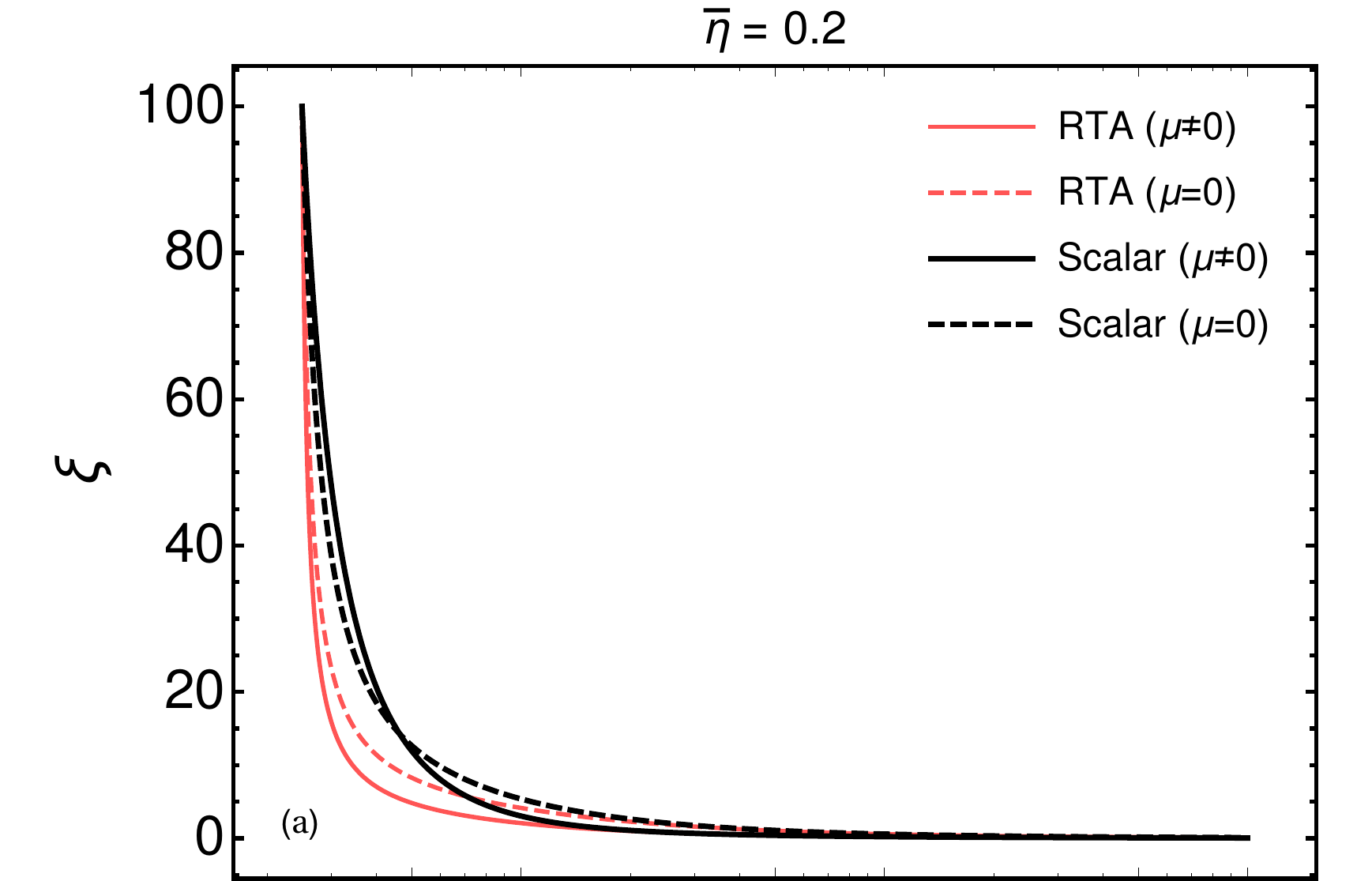}
\hspace{-7mm}
\includegraphics[width=.5\linewidth]{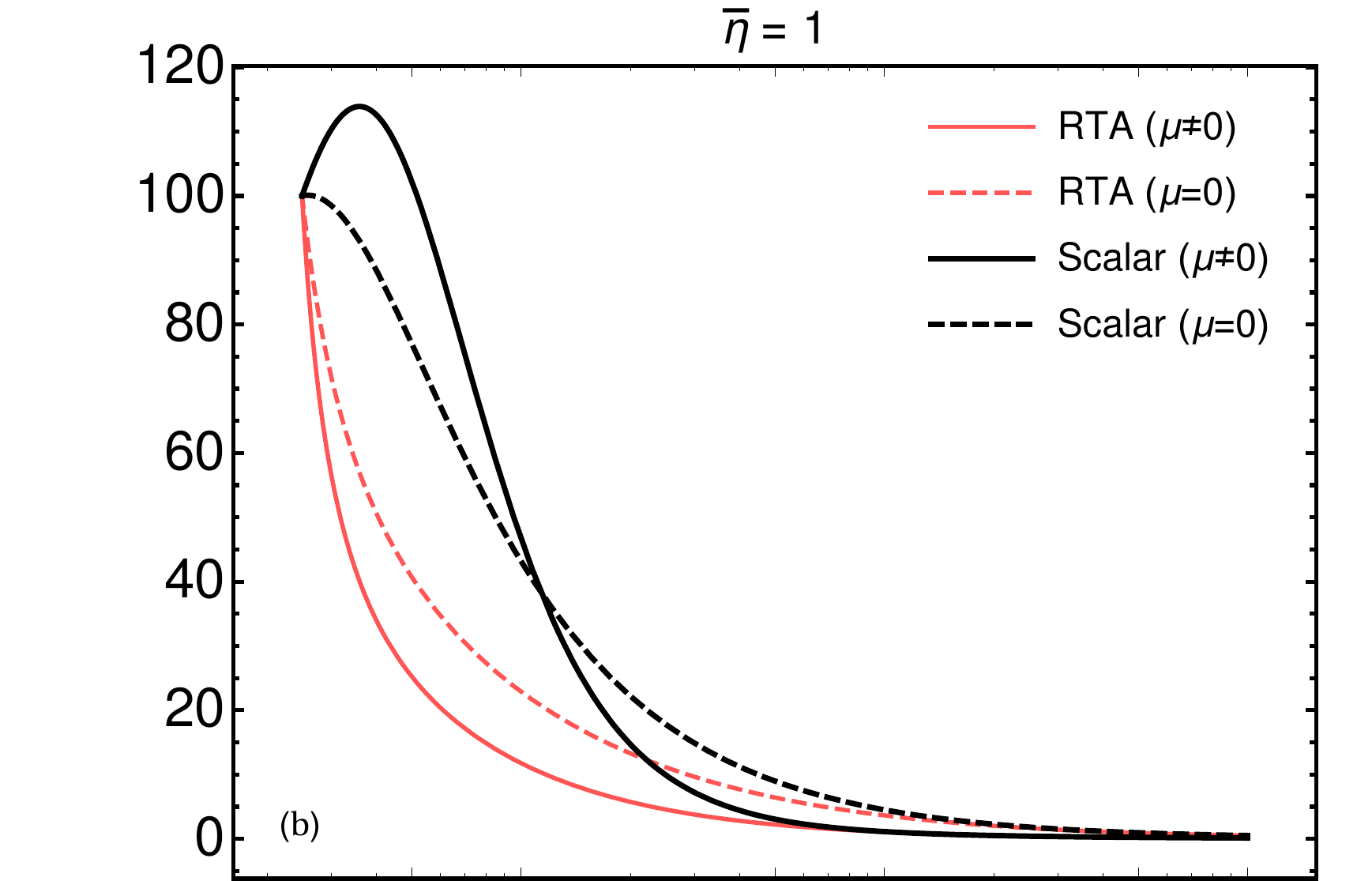}
\vspace{6mm}

\includegraphics[width=.495\linewidth]{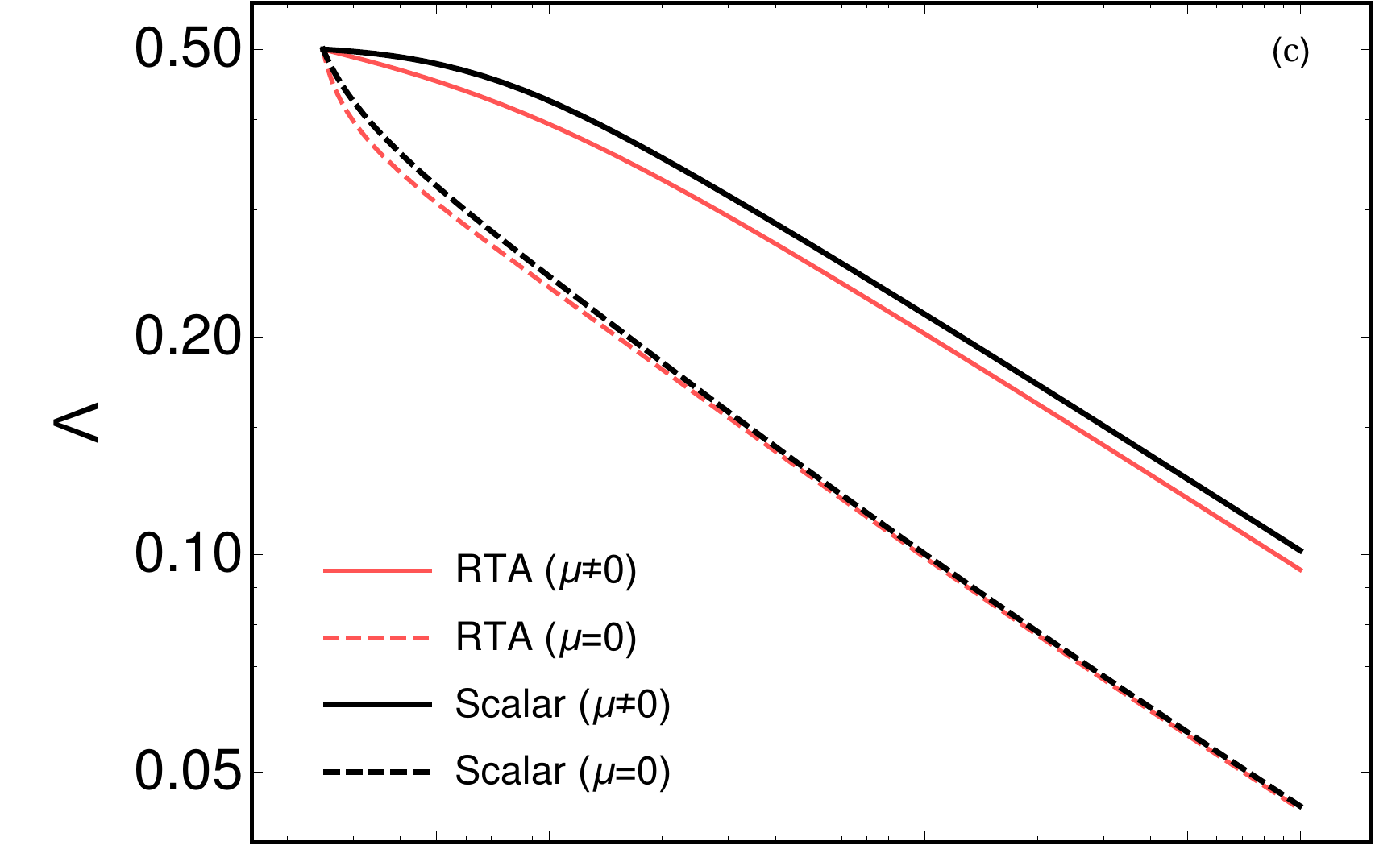}
\hspace{-7mm}
\includegraphics[width=.5\linewidth]{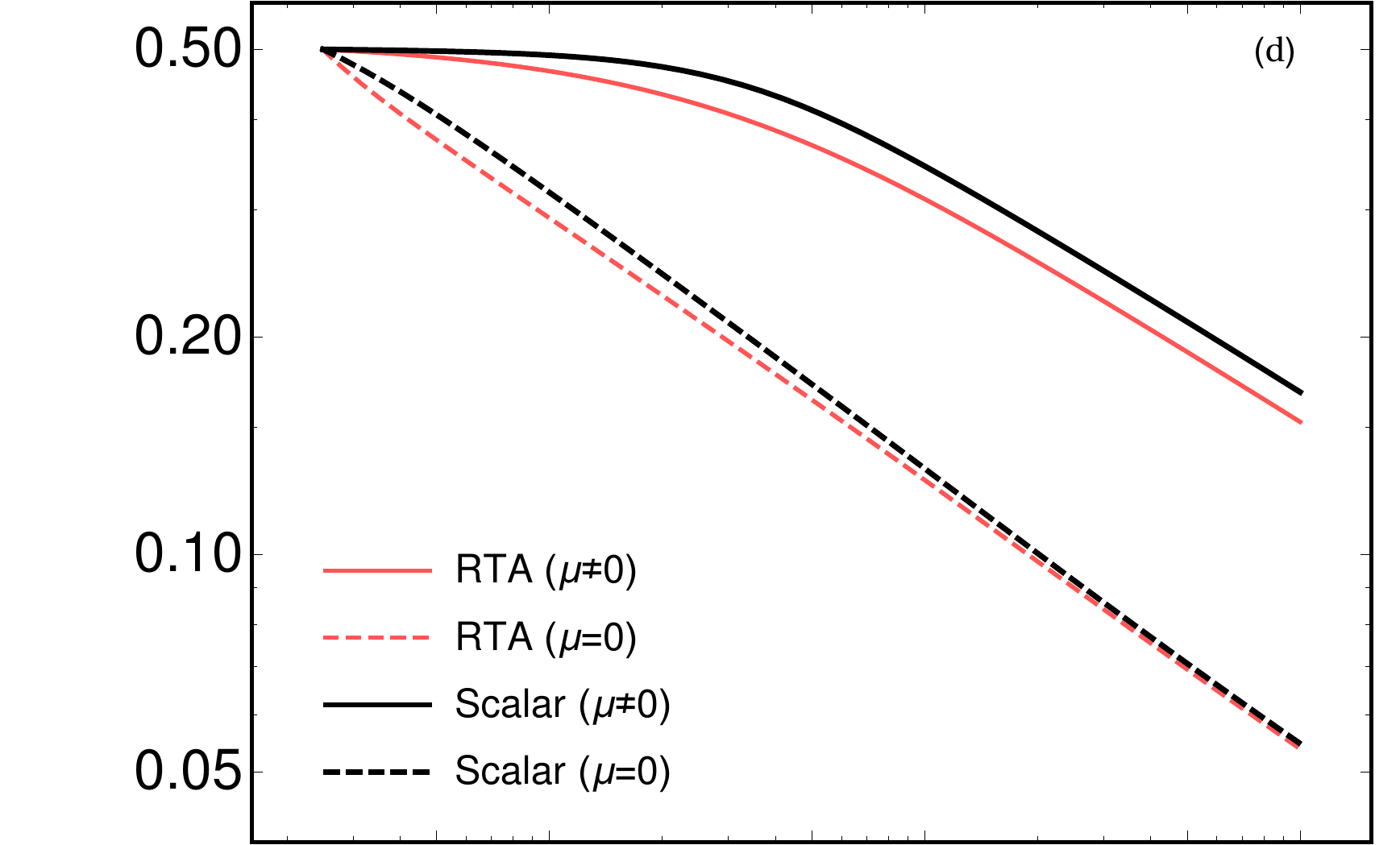}
\vspace{6mm}

\hspace{1mm}
\includegraphics[width=.46\linewidth]{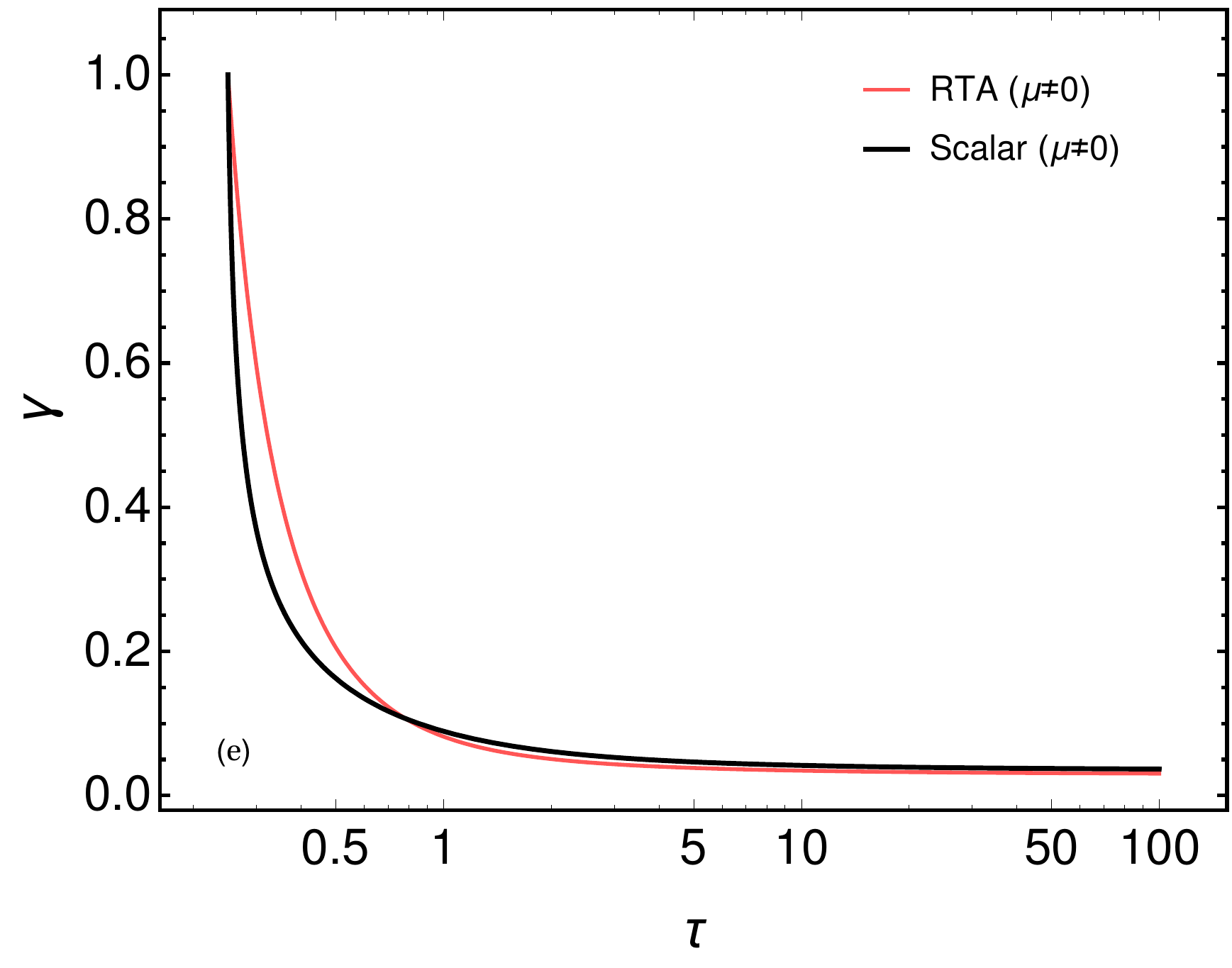}
\hspace{-1mm}
\includegraphics[width=.46\linewidth]{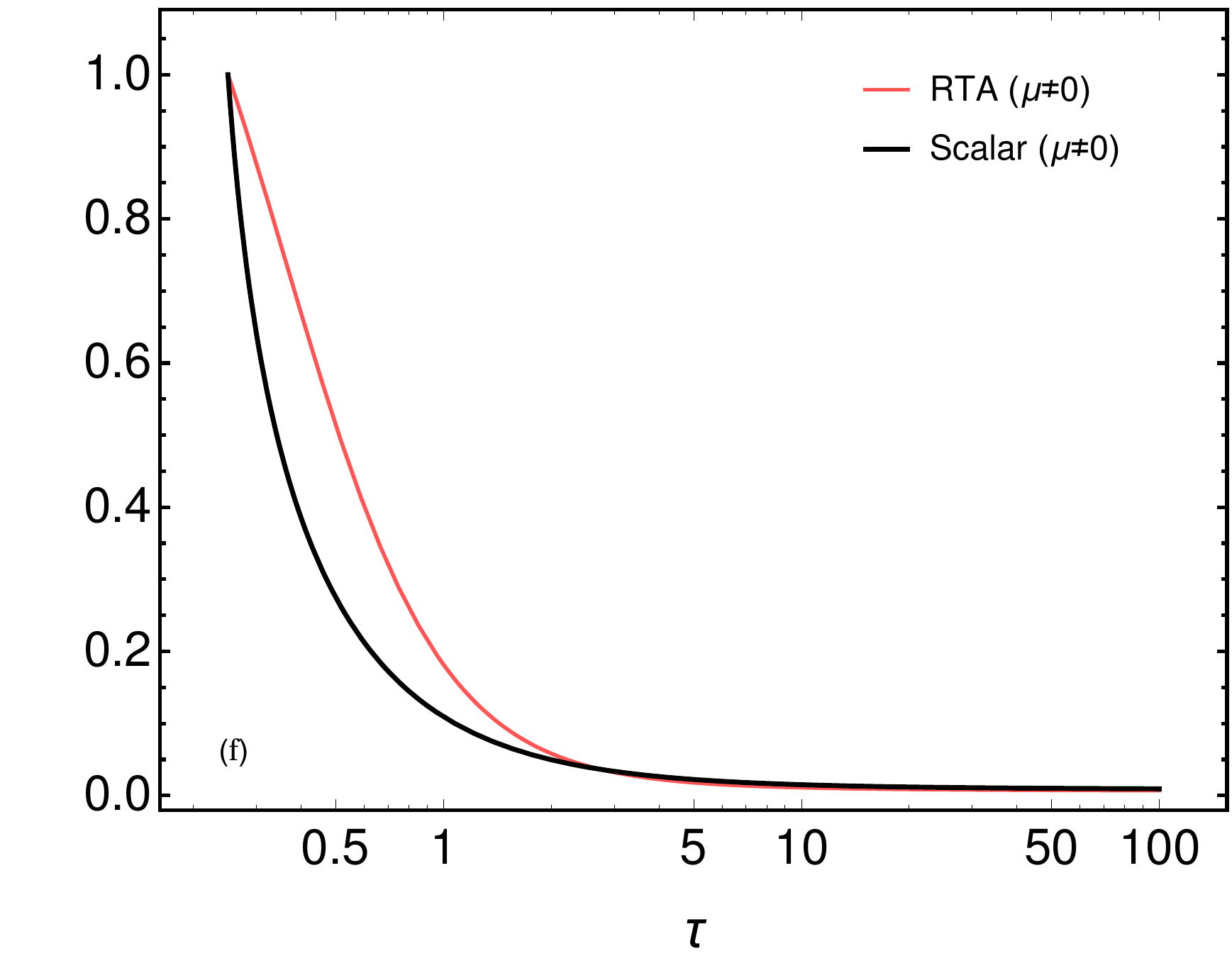}
\caption{Same as Fig.~\ref{fig:VevolPlot1} except for this figure we assumed anisotropic initial conditions  $\xi_0 = 100$}
\label{fig:VevolPlot2}
\end{figure*}

\subsection{Dynamical evolution of the microscopic parameters}

In Figs.~\ref{fig:VevolPlot1} and \ref{fig:VevolPlot2}, we present the evolution of the anisotropy paramter $\xi$, the transverse temperature scale $\Lambda$ in GeV, and the fugacity $\gamma$.  In both figures, we compare the case that $\bar\eta=0.2$ to the case when  $\bar\eta=1$.  In Fig.~\ref{fig:VevolPlot1} we assumed isotropic initial conditions with $\xi_0 = 10^{-8}$, $\tau_0= 0.25\, \rm fm/c $, $\Lambda_0 = 0.5 \,\rm GeV$, and $\gamma_0=1$.  In Fig.~\ref{fig:VevolPlot2}, we assumed anisotropic initial conditions with $\xi_0 = 100$ and all other parameters the same as Fig.~\ref{fig:VevolPlot1}.  Focussing on Fig.~\ref{fig:VevolPlot1} first, in each panel we compare the RTA and scalar collisional kernels with and without enforcing number conservation in the equations of motion.  In the top row, we see that the peak anisotropy parameter observed is consistent with the ranking hypothesized, namely that enforcing number conservation using the RTA kernel results in a reduced level of momentum-space anisotropy.\footnote{Due to the fact that we consider a conformal system with classical statistics, there is a one-to-one correspondence between the value of $\xi$ and the expected level of pressure anisotropy since the fugacity factors cancel leaving \mbox{${\cal P}_L/{\cal P}_T = {\cal R}_L(\xi)/{\cal R}_T(\xi)$} which is a monotonically decreasing function of $\xi$.}  We see the opposite ordering of the peak $\xi$ when using the scalar kernel which is consistent with our prediction that the level of momentum-space anisotropy should increase when enforcing number conservation in this case.  

Continuing on the first row of Fig.~\ref{fig:VevolPlot1}, we notice that, at late times, the RTA and scalar collisional kernels give the same asymptotic behavior, with the $\mu\neq0$ RTA and scalar results converging to one another and likewise for the case $\mu=0$.  From the second row of Fig.~\ref{fig:VevolPlot1} we see that the transverse temperature $\Lambda$ for $\mu \neq 0$ is approximately the same using either collisional kernel.  Finally, in the bottommost row of Fig.~\ref{fig:VevolPlot1} we see the evolution of the fugacity $\gamma$.  Starting from $\gamma=1$ at $\tau = 0.25$ fm/c, we see that the fugacity decreases as a function of proper time.  Turning to Fig.~\ref{fig:VevolPlot2} we observe the same patterns in the values of $\xi$ developed during the evolution.  Additionally, we see qualitatively the same behavior of the fugacity as a function of proper time, namely that it decreases monotonically and saturates to a small fixed value at late times.

\begin{figure*}[t!]
\hspace{-2mm}
\includegraphics[width=.46\linewidth]{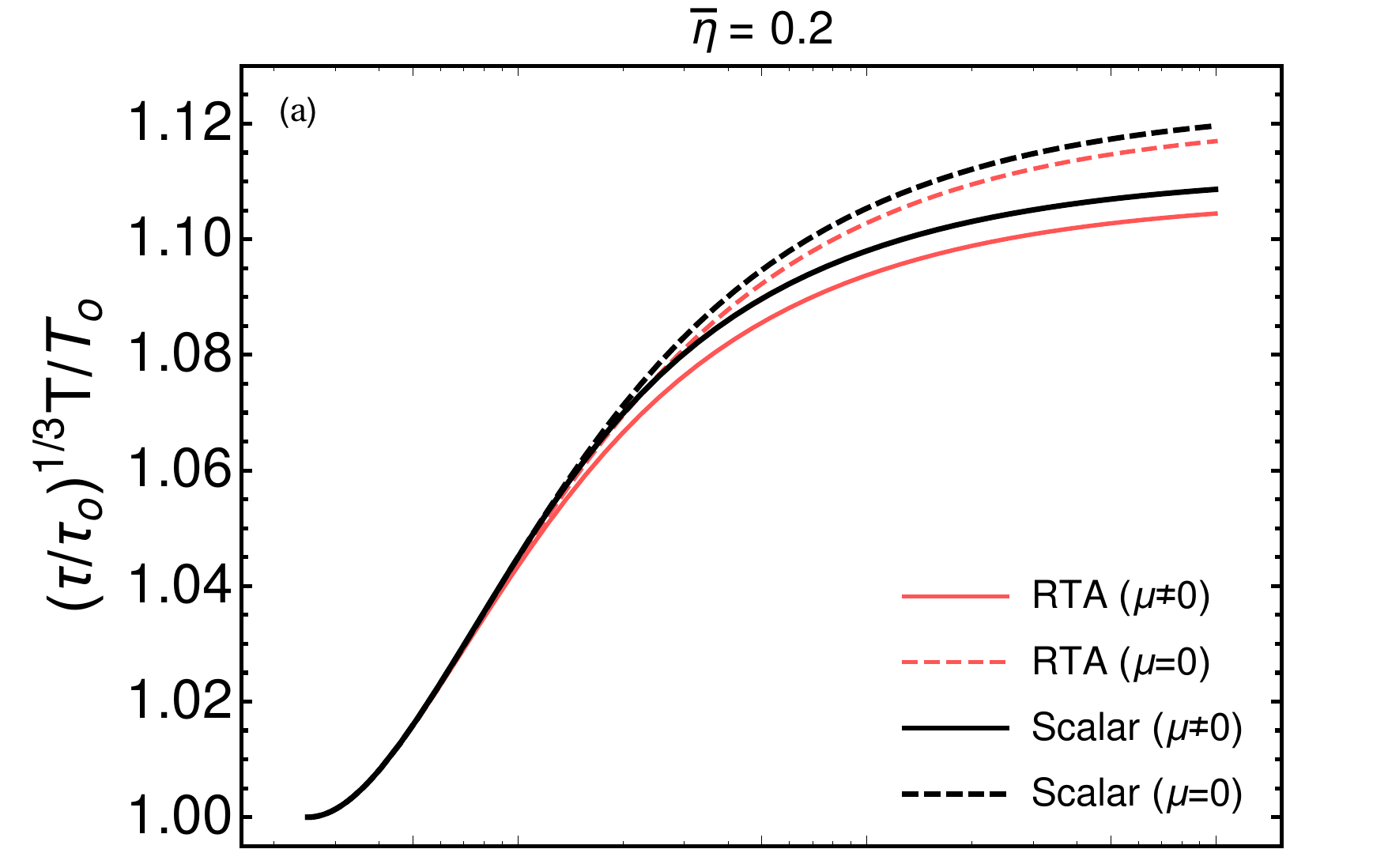}
\hspace{3mm}
\includegraphics[width=.41\linewidth]{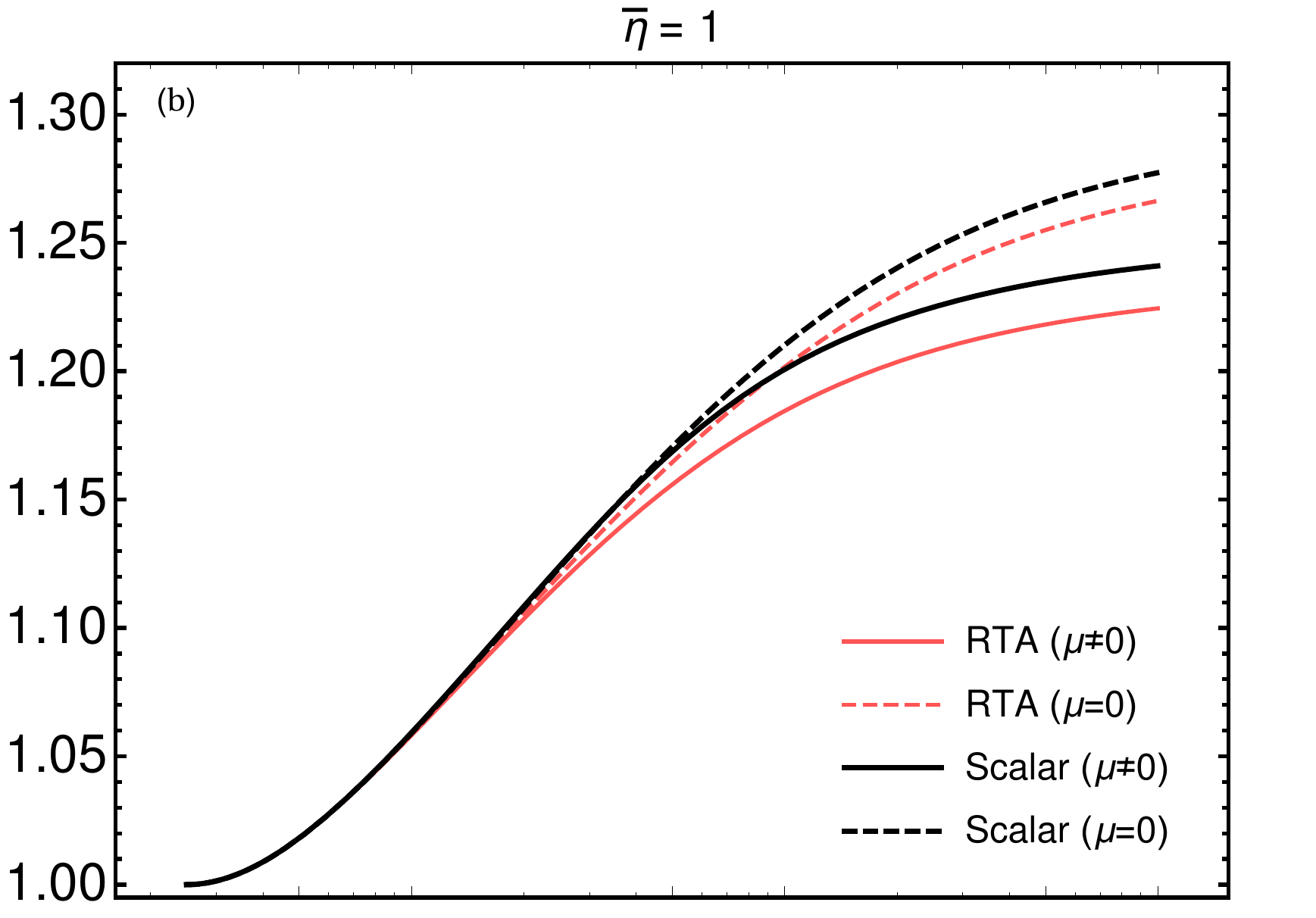}

\vspace{-4mm}
\centerline{
\hspace{6mm}
\includegraphics[width=.405\linewidth]{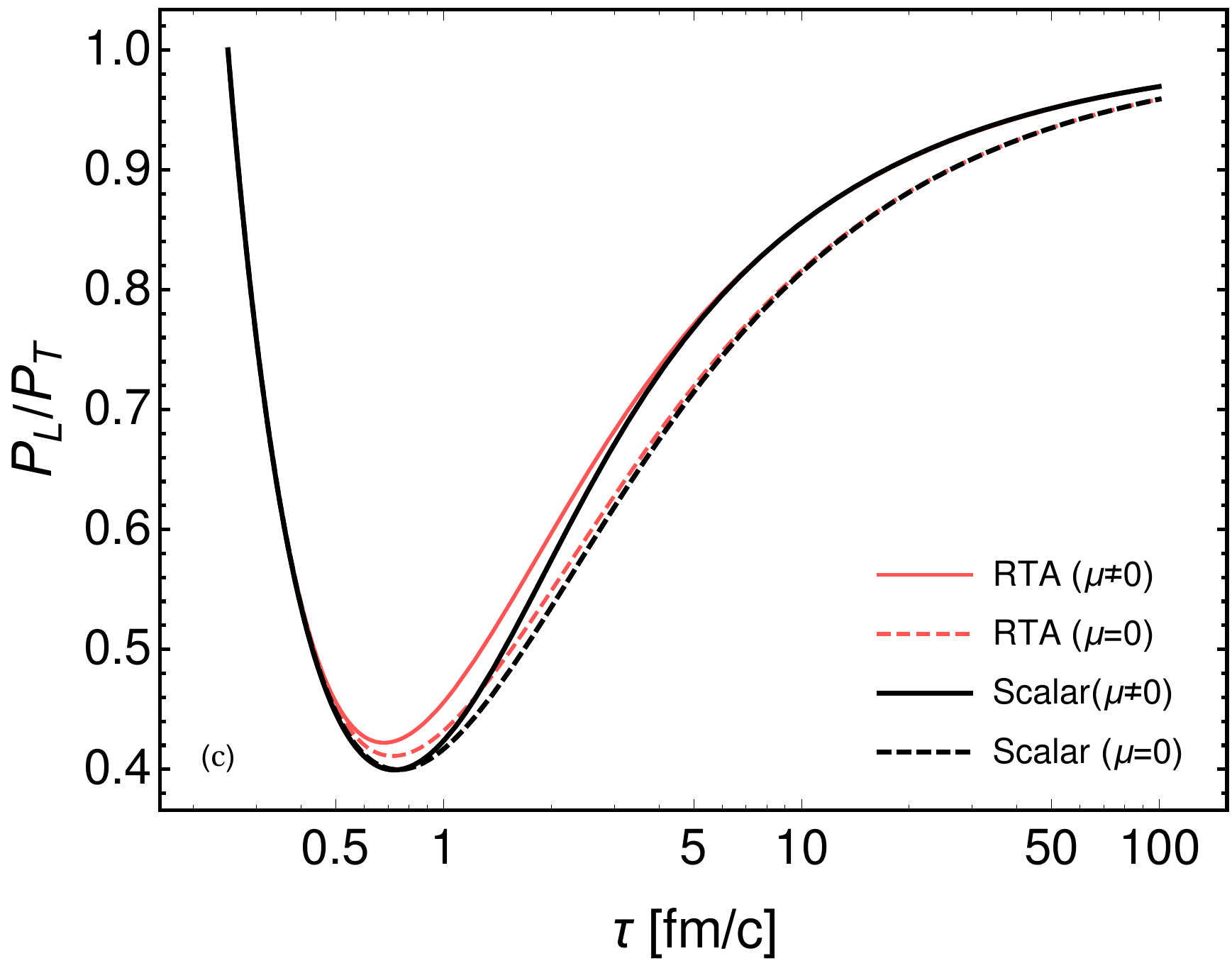}
\hspace{8mm}
\includegraphics[width=.43\linewidth]{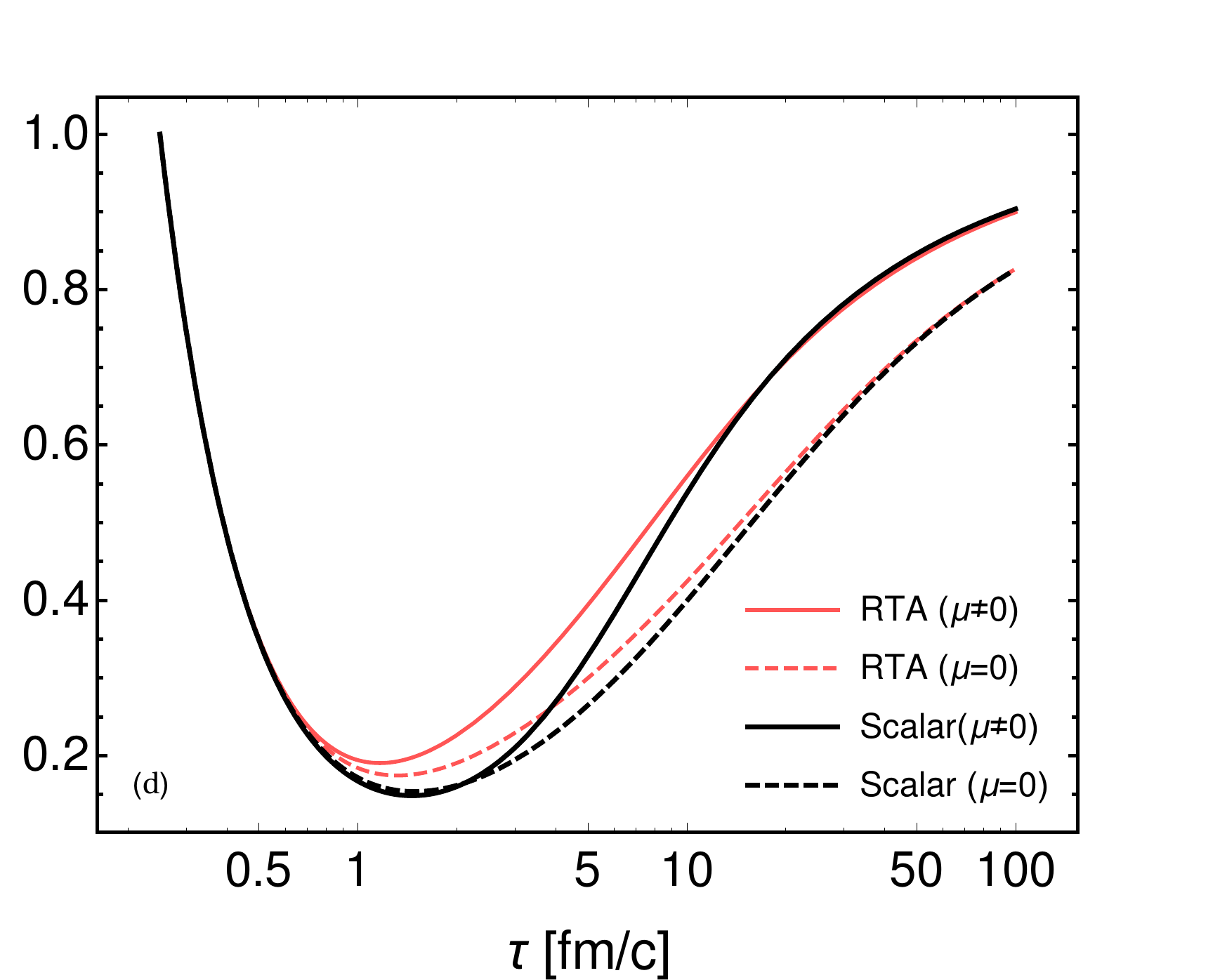}
}
\caption{Evolution of the effective temperature (a)-(b) and pressure ansiotropy (c)-(d) using both the RTA and scalar collisional kernels with and without number conservation enforced.  In the top row, we plot the scaled temperature multiplied by $(\tau/\tau_0)^{1/3}$ in order to better see the small deviations between the different approaches.  The left column panels (a) and (c) show the case $\bar\eta=0.2$ and the right column panels (b) and (d) show the case $\bar\eta=1$.  The initial conditions were taken to be isotropic with the same parameters as in Fig.~\ref{fig:VevolPlot1}.
}
\label{fig:physparms1}
\end{figure*}

\begin{figure*}[t!]
\centerline{
\hspace{-1mm}
\includegraphics[width=.47\linewidth]{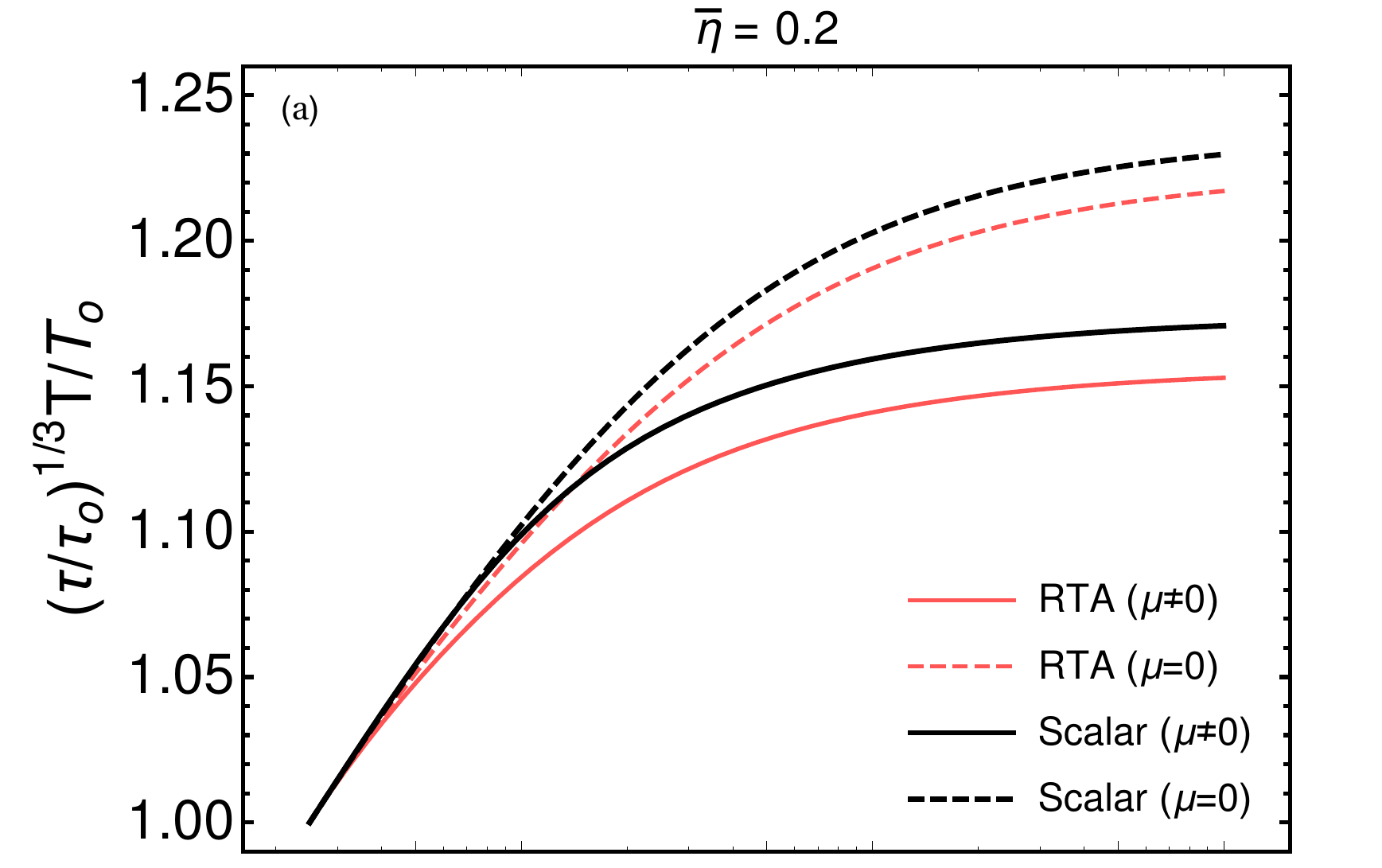}
\hspace{4mm}
\includegraphics[width=.452\linewidth]{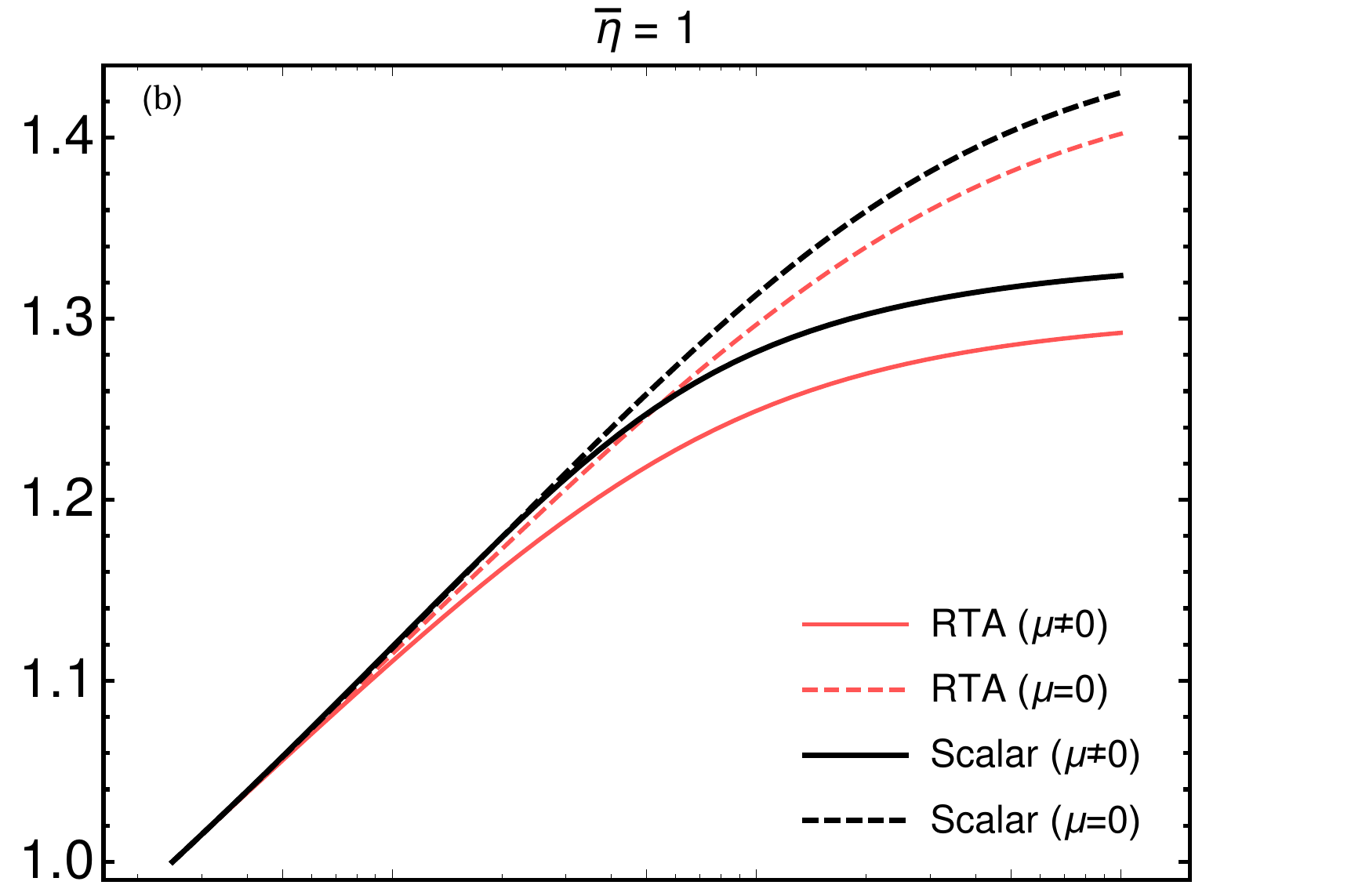}
}
\vspace{-4mm}
\centerline{
\hspace{3mm}
\includegraphics[width=.414\linewidth]{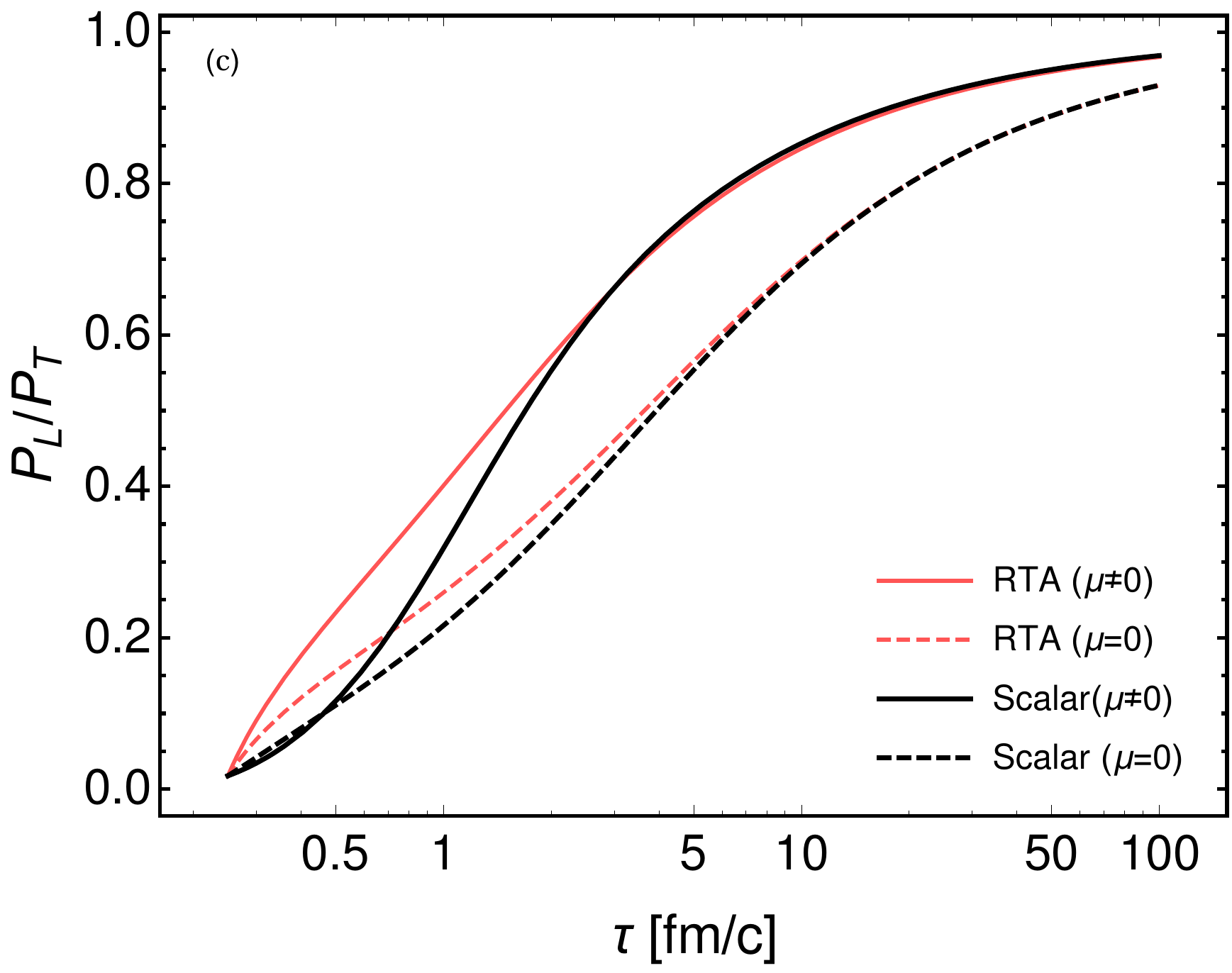}
\hspace{8mm}
\includegraphics[width=.45\linewidth]{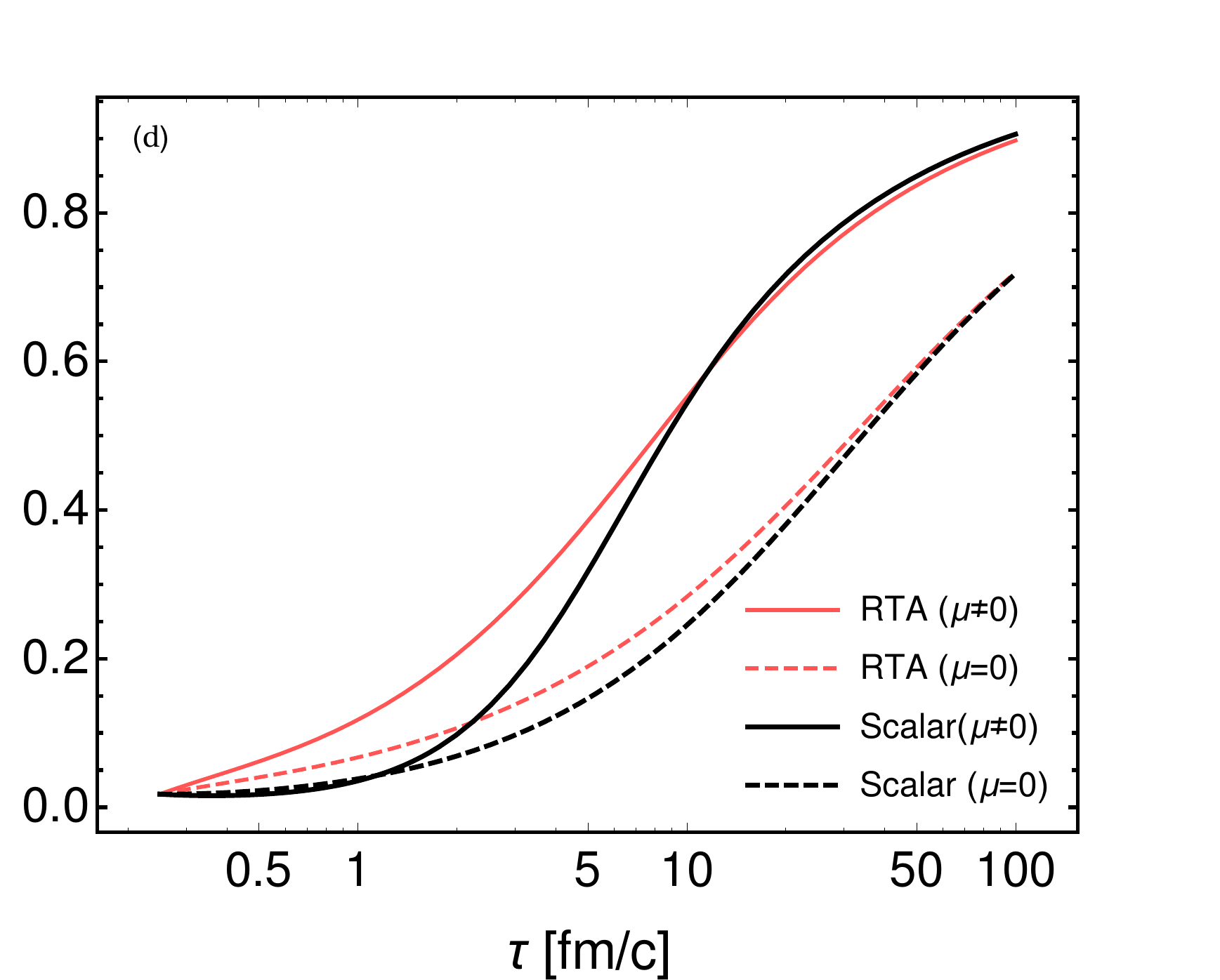}
}
\caption{Same as Fig.~\ref{fig:physparms1} except with anisotropic initial conditions.  The initial conditions and parameters are the same as in Fig.~\ref{fig:VevolPlot2}.
}
\label{fig:physparms2}
\end{figure*}

\subsection{Dynamical evolution of the effective temperature and pressure ratio}

Next we turn our attention to Figs.~\ref{fig:physparms1} and \ref{fig:physparms2} which show the effective temperature and pressure ansiotropy using both the RTA and scalar collisional kernels with and without number conservation enforced.  In Fig.~\ref{fig:physparms1} we assumed isotropic initial conditions with the same parameters as Fig.~\ref{fig:VevolPlot1}, and in Fig.~\ref{fig:physparms2} we assumed anisotropic initial conditions with the same parameters as Fig.~\ref{fig:VevolPlot2}.  In  Figs.~\ref{fig:physparms1} and \ref{fig:physparms2}, we see that both collisional kernels have the same asymptotic behavior for the pressure anisotropy for $\mu=0$ and $\mu\neq0$.  In addition, we see  only very small differences in the effective temperature which had to be multiplied by $(\tau/\tau_0)^{1/3}$ in order to make them visible to the naked eye.  At early times, we see that the ordering of the level of momentum anisotropy is consistent with our expectations based on the large-$\xi$ behaviour of the ${\cal W}$ function.  At late times, the system evolves into the small-$\xi$ region, where all collisional kernels give ${\cal W} \sim \xi$.  The late-time differences between the $\mu\neq0$ and $\mu = 0$ cases are due to the additional term involving the fugacity in the energy density evolution.  One commonality is that for both the RTA and scalar collisional kernels one sees that enforcing number conservation reduces both the late-time effective temperature and momentum-space anisotropy.

\subsection{The aHydro attractor}

Next, we turn to our numerical results for the aHydro attractor for both collisional kernels at $\mu\neq0$ and $\mu = 0$.  In both cases, given the function ${\cal W}$, one only has to solve a first order differential equation for the amplitude $\varphi$ subject to the appropriate boundary condition.  For aHydro, the boundary condition for the amplitude is \cite{Strickland:2017kux} 
\be
\lim_{\overline{w} \rightarrow 0} \varphi(\overline{w}) =  \frac{3}{4} \, .
\ee
Using this boundary condition, we then solved Eq.~(\ref{eq:ahydroattractoreq2}) numerically using built-in routines in Mathematica.

\begin{figure*}[t!]
\centerline{
\includegraphics[width=.47\linewidth]{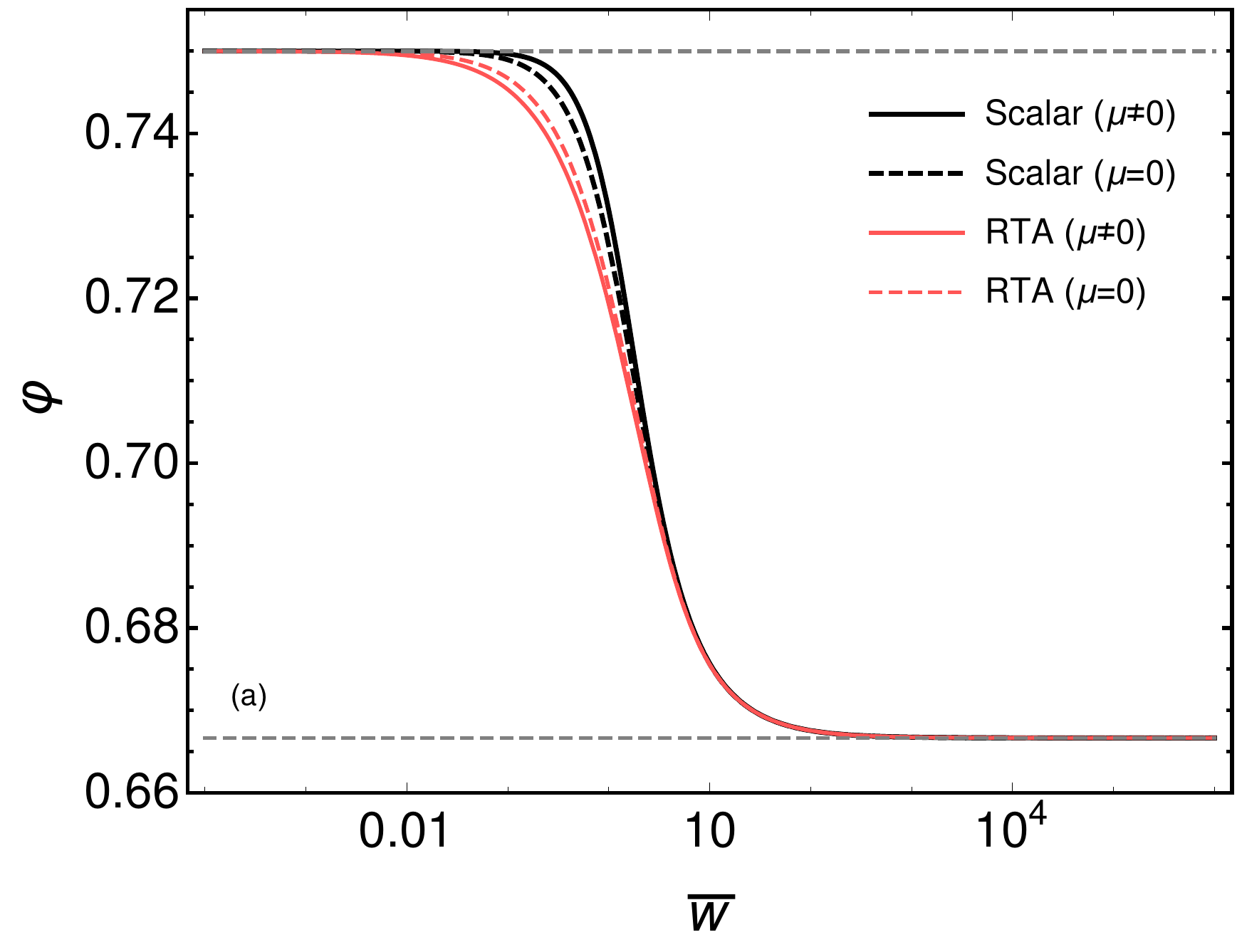}
\hspace{3mm}
\includegraphics[width=.46\linewidth]{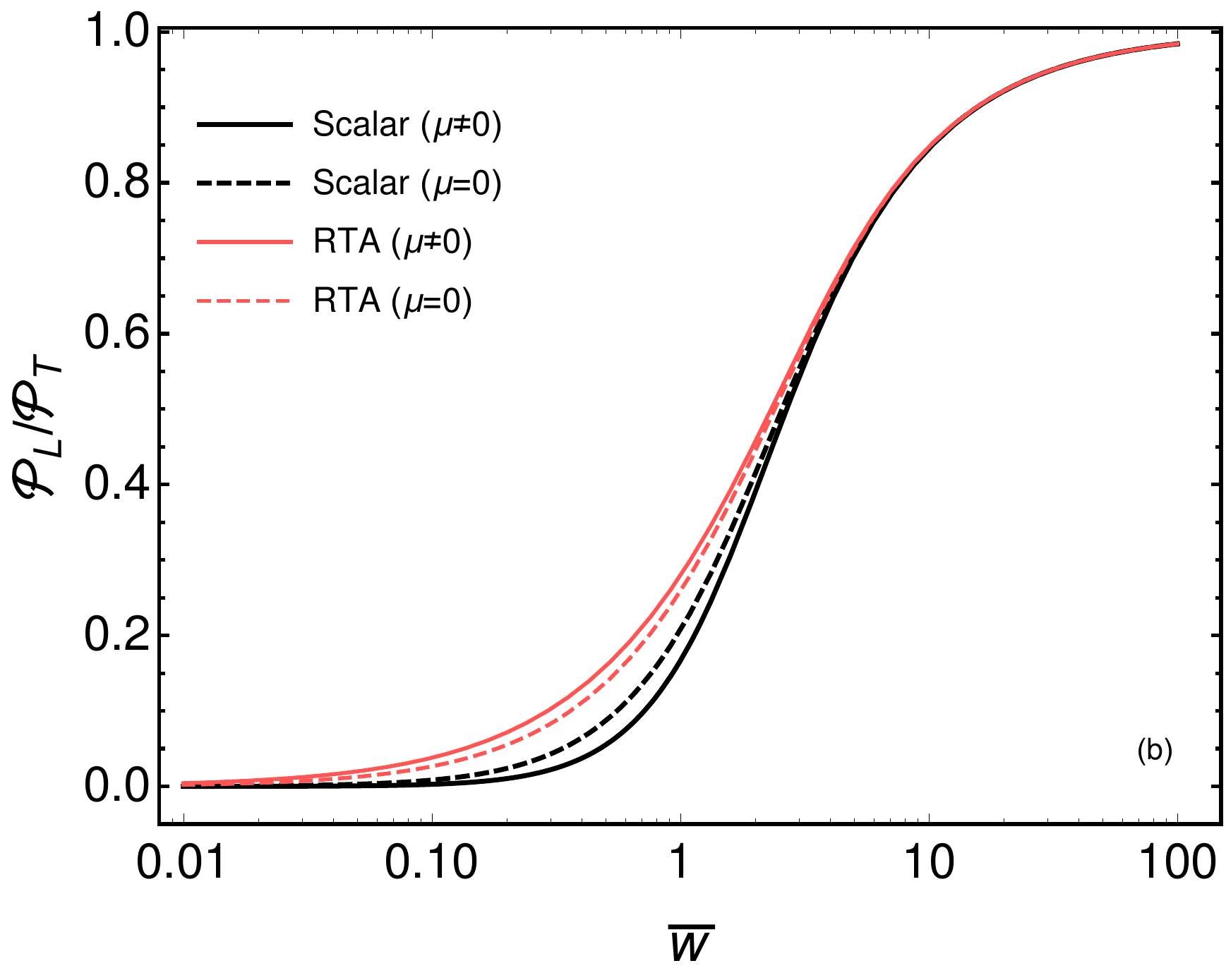}
}
\caption{The left panel (a) shows the attractor solution for the amplitude $\varphi$ and the right panel (b) shows the associated pressure anisotropy.  The four lines show results obtained using the RTA and scalar collisional kernels for both $\mu\neq0$ and $\mu = 0$.  }
\label{fig:attractor1}
\end{figure*}

In Fig.~\ref{fig:attractor1}, we compare the attractors obtained using the RTA and scalar collisional kernels for $\mu\neq0$ and $\mu = 0$.  From panel (b) we see that the effect of enforcing number conservation on the attractor is opposite when using the RTA and scalar kernels.  We see that, when we use the RTA kernel, enforcing number conservation results in less momentum-space anisotropy whereas the reverse is true for the scalar kernel.  Once again this is consistent with the observations we made in the discussion of the large-$\xi$ behavior of the $\cal W$ function.  Additionally, from this figure we see that all kernels converge to the same level of late time pressure anisotropy when plotted versus $\overline{w}$.  This rescaling gets rid of the weak dependence of the effective temperature evolution on the kernel used.

\begin{figure*}[t!]
\centerline{
\includegraphics[width=.47\linewidth]{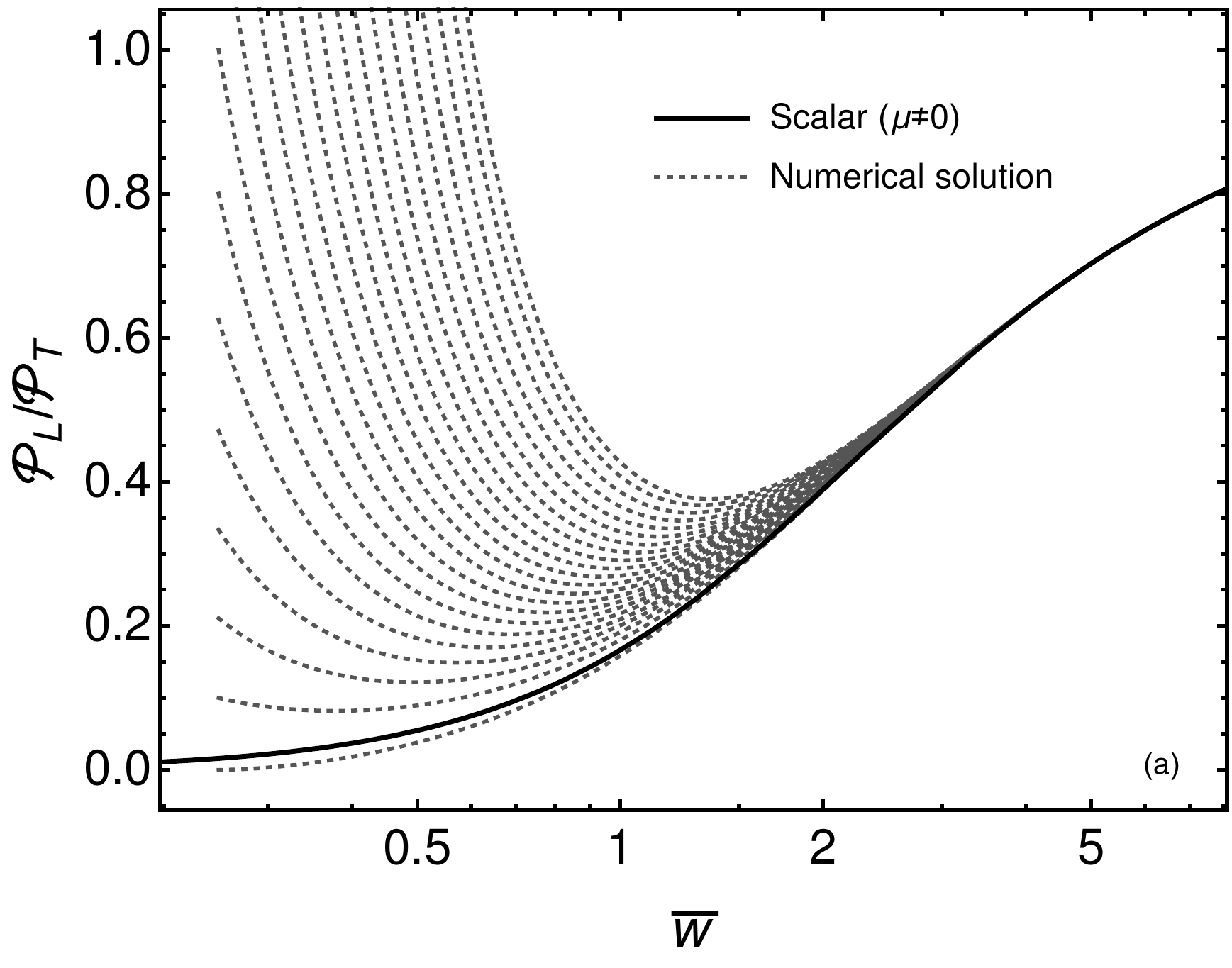}
\hspace{3mm}
\includegraphics[width=.47\linewidth]{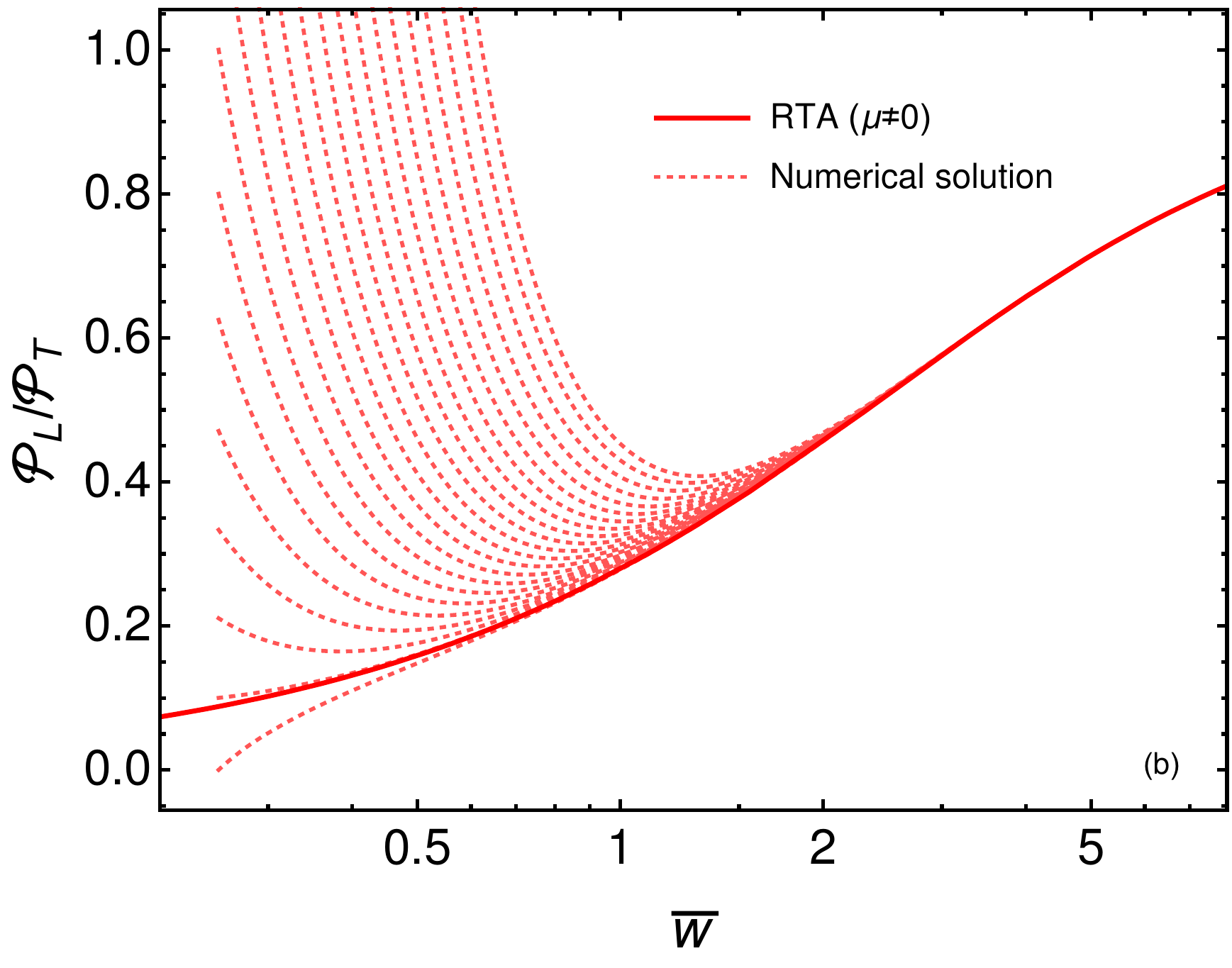}
}
\caption{Pressure anisotropy evolution for a variety of different initial conditions (dashed lines) together with the corresponding attractor (solid line).  The left panel (a) shows the results obtained using the scalar collisional kernel and the right panel (b) shows the results obtained using the RTA collision kernel.  For both panels we show the case $\mu\neq0$.
}
\label{fig:attractorcomp1}
\end{figure*}

In Fig.~\ref{fig:attractorcomp1}, we plot the pressure anisotropy evolution for a set of different initial conditions (dashed lines) together with the corresponding attractor (solid line).  The left panel (a) shows the results obtained using the scalar collisional kernel and the right panel (b) shows the results obtained using the RTA collision kernel.  For both panels we show the case $\mu\neq0$.  As can be seen from this figure, the scalar kernel results in a slightly slower rate of approach to the attractor than the RTA kernel.  This is consistent with results found in our previous paper \cite{Almaalol:2018ynx}.  Besides this, these two plots are qualitatively similar and demonstrate that one can correctly identify the attractor in aHydro when enforcing number conservation. 

\begin{figure*}[t!]
\centerline{
\includegraphics[width=.5\linewidth]{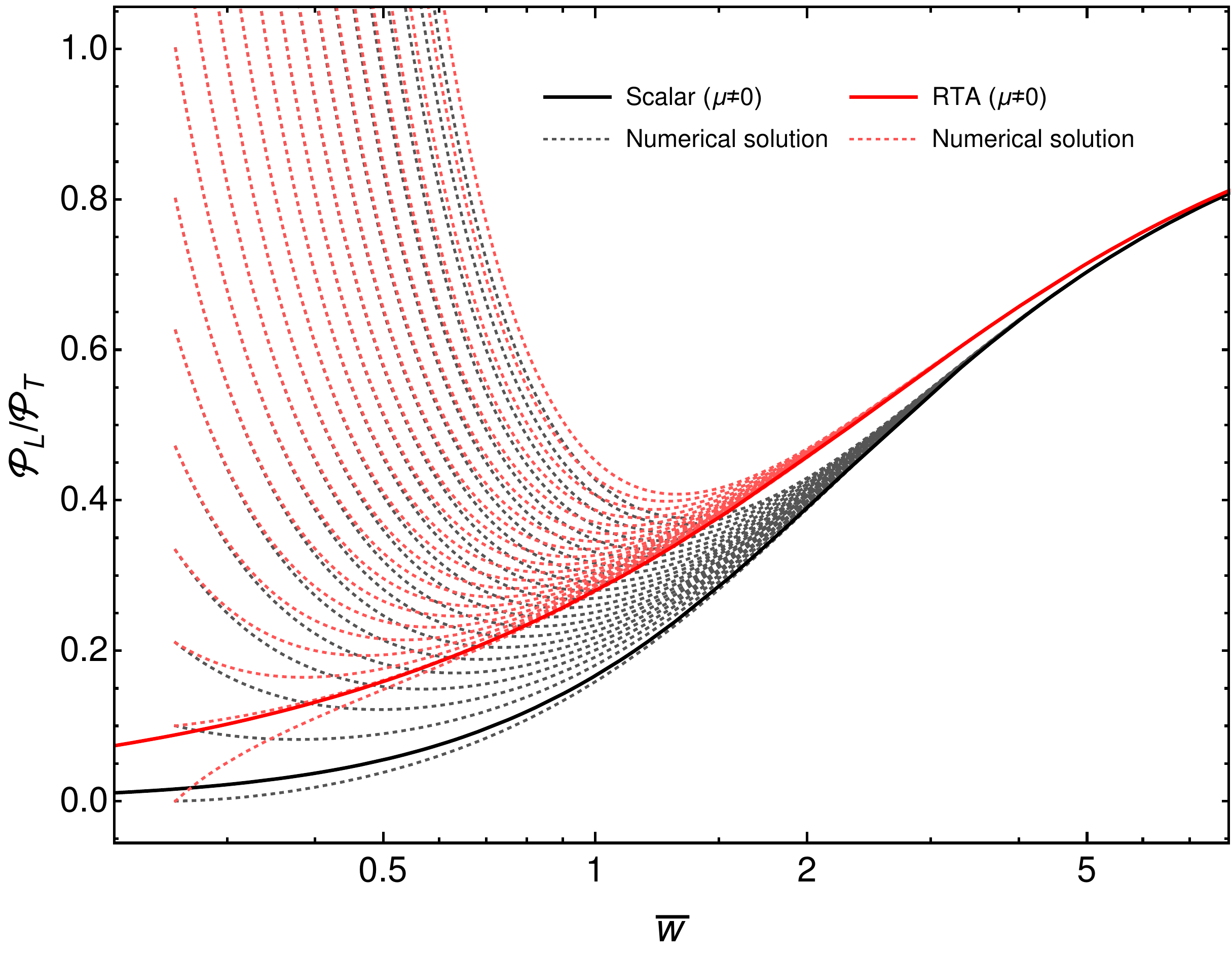}
}
\caption{Comparison of the pressure anisotropy evolution for a variety of different initial conditions (dashed lines) together with the corresponding attractor (solid lines) for both the RTA and scalar collisional kernels. }
\label{fig:attractorcomp2}
\end{figure*}

Finally, in Fig.~\ref{fig:attractorcomp2} we compare the pressure anisotropy evolution for a set of different initial conditions (dashed lines) together with the corresponding attractor (solid lines) for both the RTA and scalar collisional kernels.  As we can see clearly from this comparison, when enforcing number conservation one finds that a higher level of momentum-space anisotropy develops when using the scalar kernel than when using the RTA kernel.  Additionally, we see that, at $\overline{w} \gtrsim 5$, all results converge to a universal curve which is independent of the collisional kernel.

\section{Conclusions and outlook}
\label{sec:conclusions}

In this paper, we studied the impact of enforcing number conservation on the dynamical evolution of a 0+1d system subject to the RTA and LO conformal $\lambda \phi^4$ collisional kernels.  For both collisional kernels we obtained the necessary equations of motion for the transverse temperature $\Lambda$, anisotropy parameter $\xi$, and fugacity $\gamma$ from the first three moments of the Boltzmann equation.  For RTA, we enforced number conservation by introducing an effective fugacity $\Gamma$ in the equilibrium distribution, which was fixed using a matching condition.  For both kernels we solved the resulting coupled non-linear differential equations numerically and compared the evolution of the aHydro parameters, pressure anisotropy, and effective temperature.  

We found that, at late times, enforcing number conservation decreases both the effective temperature and pressure anisotropy for both collisional kernels considered.  At early times, however, we found a more complicated ordering of the level of pressure anisotropy when comparing the RTA and LO scalar kernels with and without enforcing number conservation.  This ordering, however, was well-explained by the behavior of the large-$\xi$ limits of each kernel's ${\cal W}$ function with $\mu=0$ and $\mu \neq 0$.  In addition to these findings, we presented the differential equation for the aHydro attractor, now taking into account number conservation.  We found that the form of the attractor equation remains the same as when not enforcing number conservation, only with a modified ${\cal W}$ function.  We solved the attractor differential equation for both collisional kernels with $\mu=0$ and $\mu \neq 0$ and compared to existing results in the literature.

The work presented herein helps us to understand the impact of different collisional kernels on aHydro evolution.  In the future, we plan to implement a realistic QCD-based collisional kernel in aHydro.  Work along these lines is in progress \cite{forth}. 

\acknowledgments

We thank H. Baza for discussions and early work on this project.  D. Almaalol was supported by a fellowship from the University of Zawia, Libya. M. Alqahtani was supported by Imam Abdulrahman Bin Faisal University, Saudi Arabia.  M. Strickland was supported by the U.S. Department of Energy, Office of Science, Office of Nuclear Physics under Award No. DE-SC0013470.

\bibliography{kernel2}

\end{document}